%% file: main.tex
\pdfoutput=1
\documentclass[12pt,a4paper]{article}

\usepackage{ifthen} 
\newboolean{pdflatex}
\setboolean{pdflatex}{true} 

\newboolean{articletitles}
\setboolean{articletitles}{true} 

\newboolean{uprightparticles}
\setboolean{uprightparticles}{false} 

\usepackage{rotating}
\usepackage{siunitx}
\newcommand{\round}[2]{\num[round-mode=places,round-precision=#1]{#2}}

\def\BpDpDmKp{$\Bp\to\Dp\Dm\Kp$\xspace}
\def\BpDzDzbKp{$\Bp\to\Dz\Dzb\Kp$\xspace}
\def\BdDzDmKp{$\Bz\to\Dz\Dm\Kp$\xspace}
\def\BDDK{$B\to D \Dbar K$\xspace}
\def\phz{\phantom{0}}
\def\phm{\phantom{$-$}}

\def\paperauthors{LHCb collaboration} 
\def\paperasciititle{Amplitude analysis of the B+ -> D+ D- K+ decay} 
\def\papertitle{Amplitude analysis\\ of the \BpDpDmKp decay} 
\def\paperkeywords{{High Energy Physics}, {LHCb}} 
\def\papercopyright{\the\year\ CERN for the benefit of the LHCb collaboration} 
\def\paperlicence{CC BY 4.0 licence}
\def\paperlicenceurl{https://creativecommons.org/licenses/by/4.0/}

\input{preamble}
\usepackage{longtable} 

\usepackage{lscape}
\begin{document}

\renewcommand{\thefootnote}{\fnsymbol{footnote}}
\setcounter{footnote}{1}

\input{title-LHCb-PAPER}


\renewcommand{\thefootnote}{\arabic{footnote}}
\setcounter{footnote}{0}

\cleardoublepage


\pagestyle{plain} 
\setcounter{page}{1}
\pagenumbering{arabic}


\input{body}

\input{acknowledgements}

\input{appendix}

\addcontentsline{toc}{section}{References}
\bibliographystyle{LHCb}
\bibliography{main,standard,LHCb-PAPER,LHCb-CONF,LHCb-DP,LHCb-TDR}

\newpage
\input{LHCb_Authorship_21-Jul-2020}

\end{document}

%% file: preamble.tex
\usepackage[top=1in, bottom=1.25in, left=1in, right=1in]{geometry}

\usepackage{microtype}
\usepackage{lineno}  
\usepackage{xspace} 
\usepackage{caption} 

\usepackage{graphicx}  
\usepackage{color}
\usepackage{colortbl}
\graphicspath{{./figs/}} 

\usepackage{amsmath} 
\usepackage{amssymb}
\usepackage{amsfonts}
\usepackage{upgreek} 

\newcommand*\patchAmsMathEnvironmentForLineno[1]{%
\expandafter\let\csname old#1\expandafter\endcsname\csname #1\endcsname
\expandafter\let\csname oldend#1\expandafter\endcsname\csname
end#1\endcsname
 \renewenvironment{#1}%
   {\linenomath\csname old#1\endcsname}%
   {\csname oldend#1\endcsname\endlinenomath}%
}
\newcommand*\patchBothAmsMathEnvironmentsForLineno[1]{%
  \patchAmsMathEnvironmentForLineno{#1}%
  \patchAmsMathEnvironmentForLineno{#1*}%
}
\AtBeginDocument{%
\patchBothAmsMathEnvironmentsForLineno{equation}%
\patchBothAmsMathEnvironmentsForLineno{align}%
\patchBothAmsMathEnvironmentsForLineno{flalign}%
\patchBothAmsMathEnvironmentsForLineno{alignat}%
\patchBothAmsMathEnvironmentsForLineno{gather}%
\patchBothAmsMathEnvironmentsForLineno{multline}%
\patchBothAmsMathEnvironmentsForLineno{eqnarray}%
}

\usepackage{hyperxmp}

\usepackage[pdftex,
            pdfauthor={\paperauthors},
            pdftitle={\paperasciititle},
            pdfkeywords={\paperkeywords},
            pdfcopyright={Copyright (C) \papercopyright},
            pdflicenseurl={\paperlicenceurl}]{hyperref}

\usepackage[colorinlistoftodos,textsize=scriptsize]{todonotes}

\usepackage[bottom,flushmargin,hang,multiple]{footmisc}

\usepackage[all]{hypcap} 

\input{lhcb-symbols-def} 

\usepackage{cite} 
\usepackage{mciteplus}

%% file: lhcb-symbols-def.tex
\usepackage{xspace} 
\usepackage{upgreek}

\def\lhcb   {\mbox{LHCb}\xspace}

\def\babar  {\mbox{BaBar}\xspace}

\def\MagUp {\mbox{\em Mag\kern -0.05em Up}\xspace}

\ifthenelse{\boolean{uprightparticles}}%
{

 \def\Pmu         {\ensuremath{\upmu}\xspace}

 \def\Ppi         {\ensuremath{\uppi}\xspace}

 \def\Ppsi        {\ensuremath{\uppsi}\xspace}

 \def\PDelta      {\ensuremath{\Delta}\xspace}                 
 \def\PXi         {\ensuremath{\Xi}\xspace}                 
 \def\PLambda     {\ensuremath{\Lambda}\xspace}                 
 \def\PSigma      {\ensuremath{\Sigma}\xspace}                 
 \def\POmega      {\ensuremath{\Omega}\xspace}                 
 \def\PUpsilon    {\ensuremath{\Upsilon}\xspace}

 \def\PB      {\ensuremath{\mathrm{B}}\xspace}                 
                  
 \def\PD      {\ensuremath{\mathrm{D}}\xspace}

 \def\PJ      {\ensuremath{\mathrm{J}}\xspace}                 
 \def\PK      {\ensuremath{\mathrm{K}}\xspace}

 \def\Pb      {\ensuremath{\mathrm{b}}\xspace}                 
 \def\Pc      {\ensuremath{\mathrm{c}}\xspace}                 
                  
 \def\Pe      {\ensuremath{\mathrm{e}}\xspace}

 \def\Pi      {\ensuremath{\mathrm{i}}\xspace}

 \def\Ps      {\ensuremath{\mathrm{s}}\xspace}

 \def\thebaroffset{0.0em}
}
{

 \def\Pmu         {\ensuremath{\mu}\xspace}

 \def\Ppi         {\ensuremath{\pi}\xspace}

 \def\Ppsi        {\ensuremath{\psi}\xspace}                 
                  
 \mathchardef\PDelta="7101
 \mathchardef\PXi="7104
 \mathchardef\PLambda="7103
 \mathchardef\PSigma="7106
 \mathchardef\POmega="710A
 \mathchardef\PUpsilon="7107
                  
 \def\PB      {\ensuremath{B}\xspace}                 
                  
 \def\PD      {\ensuremath{D}\xspace}

 \def\PJ      {\ensuremath{J}\xspace}                 
 \def\PK      {\ensuremath{K}\xspace}

 \def\Pb      {\ensuremath{b}\xspace}                 
 \def\Pc      {\ensuremath{c}\xspace}                 
                  
 \def\Pe      {\ensuremath{e}\xspace}

 \def\Pi      {\ensuremath{i}\xspace}

 \def\Ps      {\ensuremath{s}\xspace}

 \def\thebaroffset{0.18em}
}
\newcommand{\offsetoverline}[2][\thebaroffset]{\kern #1\overline{\kern -#1 #2}}%

\makeatletter
\ifcase \@ptsize \relax
  \newcommand{\miniscule}{\@setfontsize\miniscule{4}{5}}
\or
  \newcommand{\miniscule}{\@setfontsize\miniscule{5}{6}}
\or
  \newcommand{\miniscule}{\@setfontsize\miniscule{5}{6}}
\fi
\makeatother

\DeclareRobustCommand{\optbar}[1]{\shortstack{{\miniscule (\rule[.5ex]{1.25em}{.18mm})}
  \\ [-.7ex] $#1$}}

\def\epem       {{\ensuremath{\Pe^+\Pe^-}}\xspace}

\def\mumu       {{\ensuremath{\Pmu^+\Pmu^-}}\xspace}

\def\squark    {{\ensuremath{\Ps}}\xspace}

\def\cquark    {{\ensuremath{\Pc}}\xspace}

\def\bquark    {{\ensuremath{\Pb}}\xspace}

\def\pion   {{\ensuremath{\Ppi}}\xspace}

\def\pip    {{\ensuremath{\pion^+}}\xspace}
\def\pim    {{\ensuremath{\pion^-}}\xspace}

\def\kaon    {{\ensuremath{\PK}}\xspace}

\def\KorKbar {\kern \thebaroffset\optbar{\kern -\thebaroffset \PK}{}\xspace}

\def\Kp      {{\ensuremath{\kaon^+}}\xspace}
\def\Km      {{\ensuremath{\kaon^-}}\xspace}

\def\KS      {{\ensuremath{\kaon^0_{\mathrm{S}}}}\xspace}

\def\Dbar    {{\ensuremath{\offsetoverline{\PD}}}\xspace}
\def\D       {{\ensuremath{\PD}}\xspace}

\def\DorDbar {\kern \thebaroffset\optbar{\kern -\thebaroffset \PD}\xspace}
\def\Dz      {{\ensuremath{\D^0}}\xspace}
\def\Dzb     {{\ensuremath{\Dbar{}^0}}\xspace}
\def\Dp      {{\ensuremath{\D^+}}\xspace}
\def\Dm      {{\ensuremath{\D^-}}\xspace}
\def\Dpm     {{\ensuremath{\D^\pm}}\xspace}

\def\DpDm    {\ensuremath{\Dp {\kern -0.16em \Dm}}\xspace}

\def\Dstarp  {{\ensuremath{\D^{*+}}}\xspace}

\def\Ds      {{\ensuremath{\D^+_\squark}}\xspace}

\def\B       {{\ensuremath{\PB}}\xspace}

\def\BorBbar {\kern \thebaroffset\optbar{\kern -\thebaroffset \PB}\xspace}
\def\Bz      {{\ensuremath{\B^0}}\xspace}

\def\Bd      {{\ensuremath{\B^0}}\xspace}

\def\BdorBdbar {\kern \thebaroffset\optbar{\kern -\thebaroffset \Bd}\xspace}
\def\Bu      {{\ensuremath{\B^+}}\xspace}

\def\Bp      {{\ensuremath{\Bu}}\xspace}

\def\Bs      {{\ensuremath{\B^0_\squark}}\xspace}

\def\BsorBsbar {\kern \thebaroffset\optbar{\kern -\thebaroffset \Bs}\xspace}

\def\jpsi     {{\ensuremath{{\PJ\mskip -3mu/\mskip -2mu\Ppsi}}}\xspace}

\def\Y#1S{\ensuremath{\PUpsilon{(#1S)}}\xspace}

\def\LorLbar     {\kern \thebaroffset\optbar{\kern -\thebaroffset \PLambda}\xspace}

\newcommand{\decay}[2]{\ensuremath{#1\!\to #2}\xspace} 

\def\to                 {\ensuremath{\rightarrow}\xspace}

\def\AT#1     {\ensuremath{A_{\mathrm{T}}^{#1}}\xspace}           

\def\C#1      {\ensuremath{\mathcal{C}_{#1}}\xspace}                       
\def\Cp#1     {\ensuremath{\mathcal{C}_{#1}^{'}}\xspace}                    
\def\Ceff#1   {\ensuremath{\mathcal{C}_{#1}^{\mathrm{(eff)}}}\xspace}        
\def\Cpeff#1  {\ensuremath{\mathcal{C}_{#1}^{'\mathrm{(eff)}}}\xspace}       
\def\Ope#1    {\ensuremath{\mathcal{O}_{#1}}\xspace}                       
\def\Opep#1   {\ensuremath{\mathcal{O}_{#1}^{'}}\xspace}                    

\newcommand{\nospaceunit}[1]{\ensuremath{\text{#1}}}       
\newcommand{\aunit}[1]{\ensuremath{\text{\,#1}}}       

\newcommand{\tev}{\aunit{Te\kern -0.1em V}\xspace}
\newcommand{\gev}{\aunit{Ge\kern -0.1em V}\xspace}
\newcommand{\mev}{\aunit{Me\kern -0.1em V}\xspace}
\newcommand{\kev}{\aunit{ke\kern -0.1em V}\xspace}
\newcommand{\ev}{\aunit{e\kern -0.1em V}\xspace}
 
\newcommand{\mevc}{\ensuremath{\aunit{Me\kern -0.1em V\!/}c}\xspace}
\newcommand{\gevc}{\ensuremath{\aunit{Ge\kern -0.1em V\!/}c}\xspace}
\newcommand{\mevcc}{\ensuremath{\aunit{Me\kern -0.1em V\!/}c^2}\xspace}
\newcommand{\gevcc}{\ensuremath{\aunit{Ge\kern -0.1em V\!/}c^2}\xspace}
\newcommand{\gevgevcccc}{\ensuremath{\gev^2\!/c^4}\xspace} 
\newcommand{\nspmev}{\nospaceunit{Me\kern -0.1em V}\xspace}
\newcommand{\nspgev}{\nospaceunit{Ge\kern -0.1em V}\xspace}
\newcommand{\nspmevcc}{\ensuremath{\nospaceunit{Me\kern -0.1em V\!/}c^2}\xspace}
\newcommand{\nspgevcc}{\ensuremath{\nospaceunit{Ge\kern -0.1em V\!/}c^2}\xspace}
\newcommand{\nspgevgevcccc}{\ensuremath{\nspgev^2\!/c^4}\xspace} 

\def\mm   {\aunit{mm}\xspace}

\def\mum  {\ensuremath{\,\upmu\nospaceunit{m}}\xspace}

\def\fb   {\ensuremath{\aunit{fb}}\xspace}
\def\invfb   {\ensuremath{\fb^{-1}}\xspace}

\def\ps   {\ensuremath{\aunit{ps}}\xspace}

\newcommand{\chisq}{\ensuremath{\chi^2}\xspace}

\newcommand{\chisqip}{\ensuremath{\chi^2_{\text{IP}}}\xspace}

\def\gsim{{~\raise.15em\hbox{$>$}\kern-.85em
          \lower.35em\hbox{$\sim$}~}\xspace}
\def\lsim{{~\raise.15em\hbox{$<$}\kern-.85em
          \lower.35em\hbox{$\sim$}~}\xspace}

\def\pt         {\ensuremath{p_{\mathrm{T}}}\xspace}

\def\ptot       {\ensuremath{p}\xspace}

\def\evtgen     {\mbox{\textsc{EvtGen}}\xspace}

\def\geant      {\mbox{\textsc{Geant4}}\xspace}

\def\photos     {\mbox{\textsc{Photos}}\xspace}

\def\pythia     {\mbox{\textsc{Pythia}}\xspace}

\def\laura      {\mbox{\textsc{Laura}$^{++}$}\xspace}

\def\tell1  {TELL1\xspace}
\def\ukl1   {UKL1\xspace}

\newcommand{\eg}{\mbox{\itshape e.g.}\xspace}


%% file: title-LHCb-PAPER.tex

\begin{titlepage}
\pagenumbering{roman}

\vspace*{-1.5cm}
\centerline{\large EUROPEAN ORGANIZATION FOR NUCLEAR RESEARCH (CERN)}
\vspace*{1.5cm}
\noindent
\begin{tabular*}{\linewidth}{lc@{\extracolsep{\fill}}r@{\extracolsep{0pt}}}
\ifthenelse{\boolean{pdflatex}}
{\vspace*{-1.5cm}\mbox{\!\!\!\includegraphics[width=.14\textwidth]{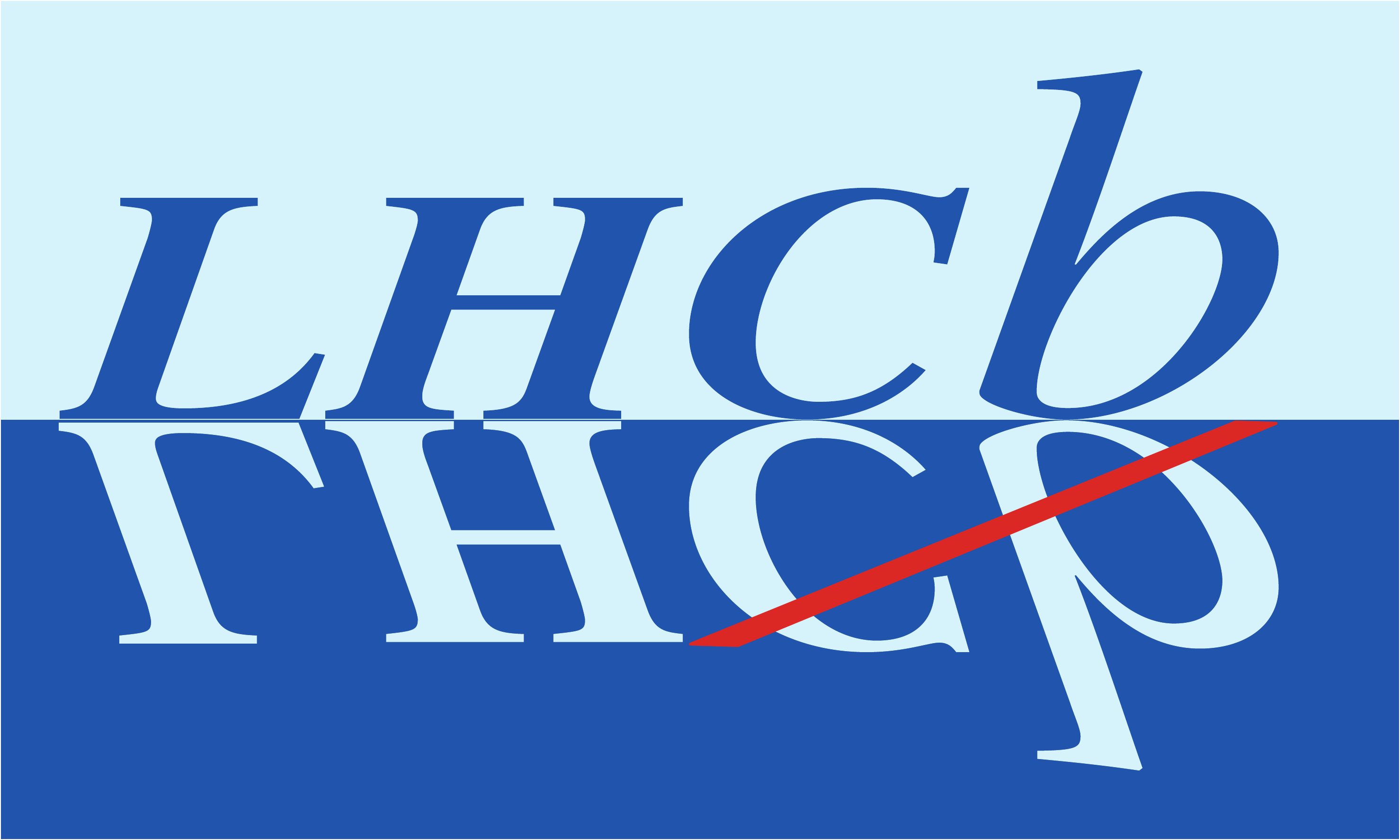}} & &}%
{\vspace*{-1.2cm}\mbox{\!\!\!\includegraphics[width=.12\textwidth]{figs/lhcb-logo.eps}} & &}%
\\
 & & CERN-EP-2020-159 \\  
 & & LHCb-PAPER-2020-025 \\  
 & & December 7, 2020 \\ 
 & & \\
\end{tabular*}

\vspace*{4.0cm}

{\normalfont\bfseries\boldmath\huge
\begin{center}
  \papertitle 
\end{center}
}

\vspace*{2.0cm}

\begin{center}
\paperauthors\footnote{Authors are listed at the end of this paper.}
\end{center}

\vspace{\fill}

\begin{abstract}
  \noindent
Results are reported from an amplitude analysis of the \BpDpDmKp\ decay. The analysis is carried out using \lhcb proton-proton collision data taken at $\sqrt{s}=7,8,$ and $13\tev$, corresponding to a total integrated luminosity of 9\invfb. In order to obtain a good description of the data, it is found to be necessary to include new spin-0 and spin-1 resonances in the $\Dm\Kp$ channel with masses around 2.9\gevcc, and a new spin-0 charmonium resonance in proximity to the spin-2 $\chi_{c2}(3930)$ state. The masses and widths of these resonances are determined, as are the relative contributions of all components in the amplitude model, which additionally include the vector charmonia $\psi(3770)$, $\psi(4040)$, $\psi(4160)$ and $\psi(4415)$ states and a nonresonant component.
  
\end{abstract}

\vspace*{2.0cm}

\begin{center}
  Published in 
  Phys.~Rev.~D 102 (2020) 112003
\end{center}

\vspace{\fill}

{\footnotesize 
\centerline{\copyright~\papercopyright. \href{\paperlicenceurl}{\paperlicence}.}}
\vspace*{2mm}

\end{titlepage}


\newpage
\setcounter{page}{2}
\mbox{~}
%

%% file: body.tex
\section{Introduction}
\label{sec:Introduction}

Decays of $B$ mesons to multibody final states involving two open-charm mesons and a strange meson, henceforth labelled \BDDK\ decays, proceed at quark level through $\bar{b} \to c\bar{c}\bar{s}$ transitions and comprise a relatively large fraction of the total width of the $B$~mesons.
Their branching fractions have been measured previously~\cite{delAmoSanchez:2010pg,Brodzicka:2007aa,LHCb-PAPER-2020-006,LHCb-PAPER-2020-015}, but few studies of their resonant structure exist.
Such analyses are valuable as a means to study resonant structure in both $D\Dbar$ and charm-strange systems.
Conventional $c\bar{c}$ charmonium states can produce resonant structures in a neutral $D\Dbar$ system, but it is now known that exotic charmonium-like states, which can decay to both neutral and charged $D\Dbar$ combinations, also exist~\cite{Karliner:2017qhf,Olsen:2017bmm,Ali:2017jda}.
Conventional resonances can also be observed in charged $DK$ systems, containing charm and antistrange ($c\bar{s}$) quarks.\footnote{The inclusion of charge-conjugate processes is implied throughout this document.}
There is no previous experimental evidence of exotic hadrons containing a charm and a strange quark ($cs$), and the possible existence of such states has not been widely discussed in the theoretical literature, although some predictions do exist~\cite{Molina:2010tx,Liu:2016ogz,Cheng:2020nho}. 

In the \BpDpDmKp\ decay, resonances in the $\Dm\Kp$ channel must have minimal quark content $\bar{c}d\bar{s}u$ and hence would be exotic, as would doubly charged $\Dp\Kp$ states. Since conventional resonances can only contribute in the $\Dp\Dm$ channel, this \B\ decay stands to provide a clean environment to study charmonium states and to address open questions concerning $c\bar{c}$ resonant structure, in particular to identify and determine the properties of spin-0 and spin-2 states~\cite{Guo:2012tv,X3915_chic02P,Duan:2020tsx}.
Properties of the vector charmonium states are better known from studies of their production in $\epem$ collisions, but improved knowledge of their rates of production in \Bp\ decays will aid characterisation of the $c\bar{c}$ contribution in $\Bp\to\Kp\mumu$ decays~\cite{Khodjamirian:2010vf,Lyon:2014hpa}.
A more detailed discussion of the current knowledge of charmomium spectroscopy, as relevant to the \BpDpDmKp decay, is given in Sec.~\ref{sec_ampmodel_content}.

No prior study of \BpDpDmKp resonant structure has been published, but a few previous amplitude analyses of other \BDDK decays exist. The Belle collaboration analysed the resonant structure of the \BpDzDzbKp decay~\cite{Brodzicka:2007aa}, while Dalitz-plot analyses of both the \BpDzDzbKp and \BdDzDmKp final states have been performed by the \babar collaboration~\cite{Lees:2014abp}.
The signal yields in these previous measurements ranged from about 400 to just under 2000, with relatively high background levels giving a maximum signal purity of 40\%.
Contributions from the vector $\psi(3770)$ and $\psi(4160)$ charmonium states, and the $D^*_{s2}(2573)^+$ and $D^*_{s1}(2700)^+$ charm-strange resonances, were determined.  
A large nonresonant contribution to the \BdDzDmKp decay was also found.

In this paper the first amplitude analysis of the \BpDpDmKp decay is described. 
The analysis is based on \lhcb proton-proton ($pp$) collision data taken at $\sqrt{s}=7,8,$ and $13\tev$, corresponding to a total integrated luminosity of 9\invfb. 
In Secs.~\ref{sec_data} and~\ref{sec_selection}, the dataset and candidate selection are described. The procedure to determine the signal and background yields, using a fit to the $\B$-candidate invariant-mass spectrum, is presented in Sec.~\ref{sec_massfit}. The amplitude modelling formalism used is detailed in Sec.~\ref{sec_amanformalism}, and a description of the selection efficiency and residual background modelling is given in Sec.~\ref{sec_eff_bgmod}. The development of the model itself follows in Sec.~\ref{sec_ampmodel}, with results given in Sec.~\ref{sec:results}. Sources of systematic uncertainties that affect the measurements are described in Sec.~\ref{sec_systematics}. Studies of the significance of various features in the model are presented in Sec.~\ref{sec_significanceTests}, and a summary of the results is provided in Sec.~\ref{sec_summary}. 

A key outcome of this amplitude analysis is the observation of structure in the $\Dm\Kp$ system.
This conclusion is confirmed with a model-independent analysis that is described in a companion article~\cite{LHCb-PAPER-2020-024}.

\section{Detector and simulation}
\label{sec_data}
The \lhcb detector~\cite{LHCb-DP-2008-001,LHCb-DP-2014-002} is a single-arm forward
spectrometer covering the \mbox{pseudorapidity} range $2<\eta <5$,
designed for the study of particles containing \bquark or \cquark
quarks. The detector includes a high-precision tracking system
consisting of a silicon-strip vertex detector surrounding the $pp$
interaction region~\cite{LHCb-DP-2014-001}, a large-area silicon-strip detector located
upstream of a dipole magnet with a bending power of about
$4{\mathrm{\,Tm}}$, and three stations of silicon-strip detectors and straw
drift tubes~\cite{LHCb-DP-2013-003,LHCb-DP-2017-001} placed downstream of the magnet.
The tracking system provides a measurement of the momentum, \ptot, of charged particles with
a relative uncertainty that varies from 0.5\% at low momentum to 1.0\% at 200\gevc.
The minimum distance of a track to a primary $pp$ collision vertex~(PV), the impact parameter (IP), 
is measured with a resolution of $(15+29/\pt)\mum$,
where \pt is the component of the momentum transverse to the beam, in\,\gevc.
Different types of charged hadrons are distinguished using information
from two ring-imaging Cherenkov detectors~\cite{LHCb-DP-2012-003}. 
Photons, electrons and hadrons are identified by a calorimeter system consisting of
scintillating-pad and preshower detectors, an electromagnetic
and a hadronic calorimeter. Muons are identified by a
system composed of alternating layers of iron and multiwire
proportional chambers~\cite{LHCb-DP-2012-002}.
The online event selection is performed by a trigger~\cite{LHCb-DP-2012-004}, 
which consists of a hardware stage, based on information from the calorimeter and muon
systems, followed by a software stage, which applies a full event
reconstruction.

At the hardware trigger stage, events are required to have a muon with high \pt or a hadron, photon or electron with high transverse energy in the calorimeters. For hadrons, the typical transverse energy threshold is 3.5\gev.
  The software trigger requires a two-, three- or four-track
  secondary vertex with a significant displacement from any primary
  $pp$ interaction vertex. At least one charged particle
  must have a transverse momentum $\pt > 1.6\gevc$ and be
  inconsistent with originating from a PV.
  A multivariate algorithm~\cite{BBDT,LHCb-PROC-2015-018} is used for
  the identification of secondary vertices consistent with the decay
  of a \bquark hadron.

Simulation is required to model the effects of the detector acceptance and the imposed selection requirements. In the simulation, $pp$ collisions are generated using \pythia~\cite{Sjostrand:2007gs,*Sjostrand:2006za} with a specific \lhcb configuration~\cite{LHCb-PROC-2010-056}. Decays of unstable particles are described by \evtgen~\cite{Lange:2001uf}, in which final-state radiation is generated using \photos~\cite{Golonka:2005pn}.   The interaction of the generated particles with the detector, and its response, are implemented using the \geant toolkit~\cite{Allison:2006ve, *Agostinelli:2002hh} as described in Ref.~\cite{LHCb-PROC-2011-006}. 
For the samples corresponding to 2017 and 2018 data, the underlying $pp$ interaction is reused multiple times, with an independently generated signal decay for each~\cite{LHCb-DP-2018-004}. 

The particle identification (PID) response in the simulated samples is corrected by sampling from distributions of $\Dstarp \to \Dz\pip, \Dz \to \Km\pip$ decays in LHCb data, considering their kinematics and the detector occupancy. An unbinned method is employed, where the probability density functions are modelled using kernel density estimation~\cite{Poluektov:2014rxa}.  The event multiplicity is also corrected in the simulated samples to match more closely that observed in events containing selected \BpDpDmKp candidates. Good agreement is seen between the simulated samples and data for the variables used in the analysis.

The momentum scale is calibrated using samples of $\decay{\jpsi}{\mumu}$ 
and $\decay{\Bu}{\jpsi\Kp}$~decays collected concurrently
with the~data sample used for this analysis~\cite{LHCb-PAPER-2012-048,LHCb-PAPER-2013-011}.
The~relative accuracy of this
procedure is estimated to be $3 \times 10^{-4}$ using samples of other
fully reconstructed $\bquark$~hadrons, $\PUpsilon$~and
$\KS$~mesons.

\section{Selection}
\label{sec_selection}

Data samples collected in $pp$ collisions during the Run~1 (2011 and 2012) and Run~2 (2015--2018) data-taking periods of the Large Hadron Collider are used, corresponding to integrated luminosities of 3\invfb and 6\invfb, respectively. Signal $\Bp$ candidates are built from sets of well-reconstructed pions and kaons, where intermediate charm mesons are reconstructed via the $\Dp\to\Km\pip\pip$ decay. 

The final-state particles are ensured to be well displaced from the interaction point by requiring that their \chisqip with respect to any PV be greater than 4, where \chisqip is defined as the difference in the vertex-fit \chisq of a given PV reconstructed with and without the particle under consideration. The PV that fits best to the flight direction of the \B candidate is taken as the associated PV. All charged final-state particles are required to have momentum greater than  $1\gevc$ and transverse momentum above $0.1 \gevc$. At least one of them must have momentum greater than $10 \gevc$ and transverse momentum exceeding $1.7 \gevc$, whilst also having an impact parameter with respect to the \B-candidate's associated PV of at least $0.1\mm$.  The \D-candidates' invariant masses are required to lie within $20$\mevcc of the known \Dpm mass~\cite{PDG2019} and their decay vertices must be well reconstructed. The reconstructed momentum and the vector between production and decay vertices are required to be well aligned for both \B and \D candidates. The flight time (distance significance) from the associated PV for the \B- (\D)-meson candidates is required to exceed $0.2 \ps$ (6). Finally, PID information is employed to aid identification of final-state \kaon and \pion mesons. 

A boosted decision tree~(BDT)~\cite{Breiman,Stevens:2013dya} algorithm, implemented in the TMVA
toolkit~\cite{Hocker:2007ht,*TMVA4}, is employed to separate signal from background. The boosting algorithm assigns weights during training both to correct for classification error and to prioritise uniformity in the Dalitz plot variables. The signal sample for the training consists of correctly reconstructed simulated \BpDpDmKp candidates and the background sample is composed of candidates from the data samples where the $\B$-candidate mass exceeds 5.6\gevcc. No evidence of overtraining is observed. Candidates are retained if the BDT response exceeds a threshold, chosen to maximise the product of signal significance and sample purity, $S^2/\left(S+B\right)^{3/2}$, where $S$ and $B$ are the expected signal and background yields in the range $5.265\gevcc < m(\Dp\Dm\Kp) < 5.295\gevcc$. The invariant mass is calculated from a kinematic fit in which the masses of the charm-meson candidates are fixed to the known $\Dpm$ mass value and the $\B$ meson is constrained to originate from its associated PV. 
Given the variations in hardware and software trigger criteria, separate BDT classifiers are developed for Run~1 and Run~2 data. The variables entering the BDT are: the $\chi^2$ of the reconstructed \B-meson decay vertex; the angle between the \B-meson flight direction from the associated PV and its reconstructed momentum; the \chisqip of the \B- and \D-meson candidates and of the final state pions and kaons; the ratio of the flight distance, parallel to the beampipe, of each of the \Dpm candidates to its uncertainty; and the PID variables of the final-state \kaon and \pion mesons. 

\begin{figure}[tb]
\centering
\includegraphics[width=.49\textwidth]{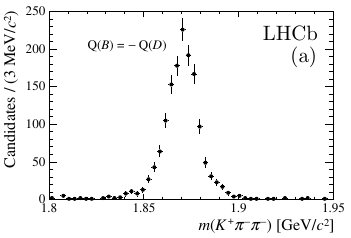}
\includegraphics[width=.49\textwidth]{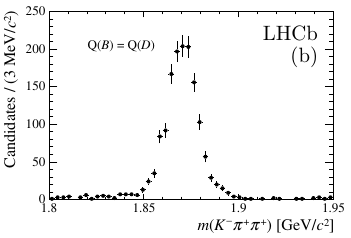}\\
\includegraphics[width=.49\textwidth]{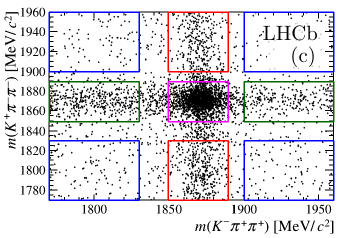}
\includegraphics[width=.49\textwidth]{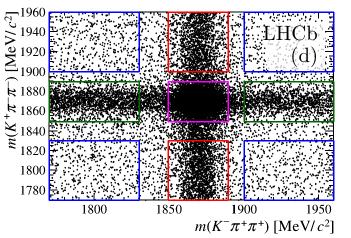}
\caption{Invariant-mass distributions for the selected candidates for the $D$ meson having (a)~the opposite and (b) the same charge, $Q$, as the $B$ meson, and in the two-dimensional plane showing the two invariant masses in (c) Run~1 and (d) Run~2 data. In (c) and (d) the blue rectangles correspond to regions of charmless background and the green and red where both single-charm and charmless processes contribute. The magenta rectangle indicates the signal region.\label{im_Dmeson_masses}}
\end{figure}

Decays of $\Bp$ mesons to the same set of final-state pions and kaons, having only one or no intermediate $\Dpm$ mesons or where final-state particles are associated to the wrong $D$ meson, are a potentially important source of background since they produce a peak in the reconstructed $\Dp\Dm\Kp$ invariant-mass distribution. 
To suppress these backgrounds, vetoes are imposed on narrow invariant-mass structures formed between specific pairs of final-state pions and kaons where the two particles originate from different $\Dpm$ mesons or the pair involves the kaon produced directly in the \B-meson decay. 
In addition, the two $\Dpm$ mesons are required to be displaced significantly, parallel to the beampipe, from their production vertex. These requirements are efficient for the \BpDpDmKp signal, and examination of the sidebands of the reconstructed $\Dpm$ invariant-mass distributions, illustrated in Fig.~\ref{im_Dmeson_masses}, confirms that there is negligible residual background contamination from this source.

The fraction of events containing more than one reconstructed candidate is measured to be below 1\%. All such candidates are retained.

\section{\boldmath{\B}-candidate invariant-mass fit}
\label{sec_massfit}

\begin{figure}[tb]
\centering
\includegraphics[width=.75\textwidth]{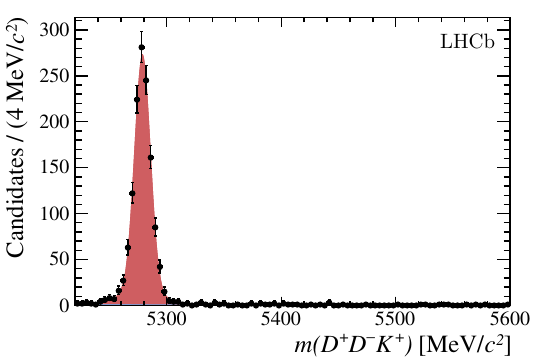}
\caption{Invariant-mass distribution for \B candidates with the results of the fit superimposed, where the signal component is indicated in red and background (barely visible) in blue.
\label{im_massfit_and_efficiencies}}
\end{figure}

An extended maximum-likelihood fit is applied to the $m(\Dp\Dm\Kp)$ distribution shown in Fig.~\ref{im_massfit_and_efficiencies}, for candidates in the range between 5.22 and 5.60\gevcc. The selected candidates in this region are predominantly from signal with a small amount of combinatorial background.
There is no significant contribution from partially reconstructed $B$ decays, which appear at lower $m(\Dp\Dm\Kp)$ values. 

The probability density function (PDF) used to model the \BpDpDmKp signal component consists of a double-sided Crystal Ball (DSCB) function~\cite{Skwarnicki:1986xj}, having tails on opposite sides of the peak in order to describe the asymmetric power-law tails of the distribution due to detector resolution and final-state radiation. An exponential function accounts for combinatorial background. In the simultaneous fit to each year of the Run~1 and Run~2 datasets, the mean and width of the signal component's Gaussian core are allowed to vary separately for the two periods, and the parameters of the DSCB tails are fixed to their values obtained in fits to simulated samples. The sample purities are very high, so if the background yield falls below 0.01 candidates for one subset of the data, the background component is removed for that subset and the fit re-run to ensure stability. The fit projection is shown in Fig.~\ref{im_massfit_and_efficiencies}, the yields of the included components are given in Table~\ref{tab_MassFitResults_yields}, and the values of the varying parameters are recorded in Table~\ref{tab_MassFitResults_shapePars}. 

Of the 1374 candidates to which the invariant-mass fit is applied, 1260 have a value of $m(\Dp\Dm\Kp)$ within 20\mevcc of the known \Bp mass, which is the window applied for the amplitude analysis. 
Within this signal window, the purity is greater than 99.5\%. 

\begin{table}[tb]
\centering
\caption{Signal and background component yields obtained from the simultaneous fit to the Run~1 and Run~2 data-taking years. \label{tab_MassFitResults_yields}}
\begin{tabular}{l @{\hspace{5mm}} c @{\hspace{5mm}} c}
\hline
Year & Signal  & Background \\
\hline
2011 & $\phz84 \pm \phz9$ & --- \\
2012 & $217 \pm 15$ & $16 \pm 5$\\
2015 & $\phz41 \pm \phz6$ & --- \\
2016 & $300 \pm 18$ & $19 \pm 6$\\
2017 & $302 \pm 18$ & $21 \pm 6$\\
2018 & $359 \pm 19$ & $15 \pm 5$\\
\hline
\end{tabular}
\end{table}

\begin{table}[tb]
\centering
\caption{Fitted values of shape parameters of the DSCB and exponential PDFs used to model signal and background, respectively, in the simultaneous fit to Run~1 and Run~2 data. \label{tab_MassFitResults_shapePars}}
\begin{tabular}{l l r @{$~\pm~$} l}
\hline
\multicolumn{2}{l}{Parameter} & \multicolumn{2}{c}{Result}  \\
\hline
{Signal}\\
$\mu$ (\nspmevcc) & Run~1  & $5278.90$ & $0.39$\\
 & Run~2  & $5278.70$ & $0.27$\\
$\sigma$ (\nspmevcc) & Run~1  & $6.22$ & $0.33$\\
 & Run~2  & $7.77$ & $0.23$\\
 \hline
{Background}\\
Coefficient (10\gevcc)$^{-1}$  & 2012 & $-38$  & $31$ \\
 & 2016 & $-93$  & $31$ \\
 & 2017 & $-66$  & $28$ \\
 & 2018 & $2$  & $36$ \\                        
\hline
\end{tabular}
\end{table}

\section{Amplitude analysis formalism}
\label{sec_amanformalism}

The distribution of \BpDpDmKp\ decays across the Dalitz plot is fitted using the \laura\ software package~\cite{Back:2017zqt}.
Generic details of the formalism and its implementation in the analysis of LHCb data can be found in the literature~\cite{LHCb-PAPER-2014-036,LHCb-PAPER-2016-026,LHCb-PAPER-2019-017}; only aspects specifically relevant to the current analysis are described here.

The PDF used to fit the Dalitz plot structure of the selected candidates is composed of signal and background contributions and is a function of position in the $\B$-decay phase space, $\vec{x}$. It includes dependence upon model parameters such as mass, width, or spin of individual components in the signal model. The fit procedure maximises the likelihood, 

\begin{equation}
    \mathcal{L} = \exp{\left[-\sum_c \left(\frac{(p_c-\mu_c)^2}{2\sigma_c^2}\right)\right]} \prod_{j=1}^{N_c} \left( N_{\rm sig} \mathcal{P}_{\rm sig}(\vec{x}_j) + N_{\rm bg} \mathcal{P}_{\rm bg}(\vec{x}_j) \right)\,, 
\end{equation}
where $N_{\rm sig}$ and $N_{\rm bg}$ are the signal and background yields obtained from the invariant-mass fit, respectively, $N_c$ is the total number of candidates in the data sample, and $\mathcal{P}_{\rm sig, bg}(\vec{x}_j)$ are the PDFs for candidate $j$, which differ for Run~1 and Run~2 data since different efficiency and background models are employed. Gaussian constraints with parameters $\mu_c$ and $\sigma_c$ are applied to the values of model parameters, $p_c$, such as the masses or widths of intermediate resonances given in Sec.~\ref{sec_ampmodel}.
The background PDF, $\mathcal{P}_{\rm bg}(\vec{x})$, is an empirical shape used to represent the residual combinatorial background that enters the selected sample of \BpDpDmKp candidates, and is described further in Sec.~\ref{sec_eff_bgmod}. 

The signal PDF is given by
\begin{equation}
    \mathcal{P}_{\rm sig}(\vec{x}) = \frac{1}{{\cal N}} \times \epsilon_{\rm total}(\vec{x}) \times |\mathcal{A}_{\rm sig}(\vec{x})|^2\,,
\end{equation}
where ${\cal N}$ is a normalisation factor that ensures the integral of $\mathcal{P}_{\rm sig}(\vec{x})$ over the Dalitz plot ($\vec{x}$) is unity, and $\epsilon_{\rm total}(\vec{x})$ is the total efficiency for the selected candidates, described further in Sec.~\ref{sec_eff_bgmod}. The signal amplitude, $\mathcal{A}_{\rm sig}(\vec{x})$, is constructed according to the isobar formalism~\cite{Fleming:1964zz,Morgan:1968zza,Herndon:1973yn} and contains a coherent sum of resonant and nonresonant amplitudes,
\begin{equation}
    \mathcal{A}_{\rm sig}(\vec{x}) = \sum_{j=1}^{N} c_j F_j(\vec{x})\,, \label{eqn_isobarsum}
\end{equation}
where the sum runs over the components in the model, indexed by $j$. The $c_j$ factors are complex coefficients that multiply the complex amplitudes $F_j(\vec{x})$, which contain information about the dynamics of each component in the amplitude model. For a \Dp\Dm resonance, for example,
\begin{equation}
    F(\vec{x}) = R\left(m(\Dp\Dm)\right)\times T(\vec{p},\vec{q}) \times X(|\vec{p}| ) \times X(|\vec{q}|)\,,
\end{equation}
where $R$ and $T$ describe the invariant-mass and angular dependence of the amplitude, and the $X$ functions are Blatt--Weisskopf barrier factors. The invariant-mass dependence, $R\left(m(\Dp\Dm)\right)$, is given by a relativistic Breit--Wigner function for all resonant contributions and the angular terms, $T(\vec{p},\vec{q})$, are constructed using the non-relativistic Zemach tensor formalism~\cite{Zemach:1963bc,Zemach:1968zz}.
Nonresonant contributions are described with a lineshape that includes an exponential form factor, with alternative models also considered during the model-building and determination of systematic uncertainties.
The momenta $\vec{p}$ and $\vec{q}$ are those of the third particle (not involved in the resonance) and one of the particles produced in the resonance decay, respectively, both evaluated in the rest frame of the resonance.

The choice of which of the particles produced in the resonance decay is taken to define $\vec{q}$ corresponds to a convention for the definition of the helicity angle of the resonance.
The helicity angle is defined to be, in the rest frame of the resonance, the angle between one of the two particles produced in the resonance decay and the third particle. 
In this study, the choice is:
\begin{itemize}
\item $\theta(\Dp\Dm)$ is the angle between the \Kp and \Dm particles, in the \Dp\Dm rest frame,
\item $\theta(\Dp\Kp)$ is the angle between the \Dm and \Kp particles, in the \Dp\Kp rest frame, and
\item $\theta(\Dm\Kp)$ is the angle between the particles \Dp and \Kp, in the \Dm\Kp rest frame.
\end{itemize}

The square Dalitz plot (SDP) provides a useful representation of the phase space. The large \Bp mass means that resonant structure is often found close to the edge of the regular Dalitz plot, and the SDP provides greater granularity in exactly these regions. Moreover, the SDP aligns a rectangular grid with the edges of the phase space, avoiding edge effects associated with rectangular binning of the regular Dalitz plot. 

The two degrees of freedom used to define the SDP are the variables $m^\prime\left(\Dp\Dm\right)$ and $\theta^\prime\left(\Dp\Dm\right)$, which are defined as
\begin{eqnarray}
m^\prime\left(\Dp\Dm\right) &\equiv& \frac{1}{\pi} \arccos{\left( 2\;\frac{ m(\Dp\Dm) - m_{\Dp\Dm}^{\rm min} }{ m_{\Dp\Dm}^{\rm max} - m_{\Dp\Dm}^{\rm min} } - 1 \right)}\,,\\
\theta^\prime\left(\Dp\Dm\right) &\equiv& \frac{1}{\pi} \theta(\Dp\Dm)\,,
\end{eqnarray}
where $m_{\Dp\Dm}^{\rm min,max}$ are the minimum and maximum kinematically allowed values of $m(\Dp\Dm)$  (equal to $2m_{\Dp}$ and $m_{\Bp} - m_{\Kp}$, respectively). 
With these definitions both $m^\prime$ and $\theta^\prime$ are bounded in the range 0--1.

The complex coefficients, $c_j$ in Eq.~\eqref{eqn_isobarsum}, depend on choices of phase convention and normalisation. In order to be able to compare results between different analyses, it is therefore helpful to report the convention-independent fit fractions, which are defined as the integral of the absolute value of the amplitude squared for each component, $j$, divided by that of the coherent matrix-element squared for all intermediate contributions,
\begin{equation}
    \mathcal{F}_j = \frac{ \int |c_j F_j(\vec{x})|^2 d\vec{x} }{ \int |\mathcal{A}_{\rm sig}(\vec{x})|^2 d\vec{x} }\,.
\end{equation}
Interference between amplitudes in the coherent sum within $\mathcal{A}_{\rm sig}(\vec{x})$ can cause the sum of the fit fractions to depart from unity. This deviation can be quantified by means of interference fit fractions,
\begin{equation}
    \mathcal{I}_{ij} = \frac{ \int c_i c_j^* F_i(\vec{x}) F_j^*(\vec{x}) d\vec{x} }{ \int |\mathcal{A}_{\rm sig}(\vec{x})|^2 d\vec{x} }\,.
\end{equation}
Interference effects between different partial waves in the same two-body combination cancel when integrated over the helicity angle, due to the angular terms having the form of Legendre polynomials, which form an orthogonal basis.

\section{Efficiency and background models}
\label{sec_eff_bgmod}
\begin{figure}[!tb]
\centering
\includegraphics[width=.49\textwidth]{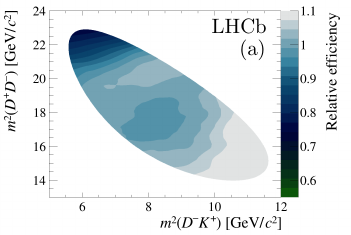}
\includegraphics[width=.49\textwidth]{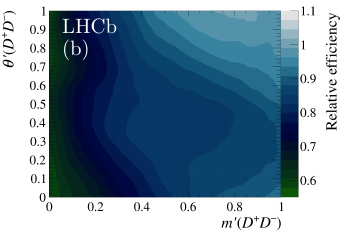}\\
\includegraphics[width=.49\textwidth]{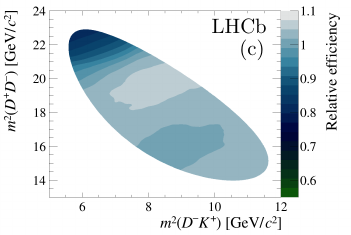}
\includegraphics[width=.49\textwidth]{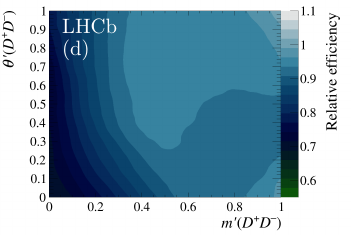}
\caption{Efficiency maps for (upper) Run~1  and (lower) Run~2, where the variation as a function of position in the  (left) standard Dalitz plot and (right) SDP are shown. 
The $z$-axis scale is arbitrary as the absolute efficiency does not affect the analysis. \label{im_efficiencyVariation}}
\end{figure}

The absolute efficiency is not needed for the amplitude analysis but the variation of the efficiency across the Dalitz plot must be accounted for. 
Efficiency variations as a function of position in the Dalitz plot are evaluated using simulated samples. Four contributing factors are considered:
\begin{equation}
    \epsilon_{\rm total}(\vec{x}) = \epsilon_{\rm offline}(\vec{x}) \times \epsilon_{\rm reco}(\vec{x}) \times \epsilon_{\rm trig}(\vec{x}) \times \epsilon_{\rm geom}(\vec{x})\,.
    \label{eqn_effs}
\end{equation}
The geometrical efficiency, $\epsilon_{\rm geom}$, quantifies the probability for all final state particles to be within the \lhcb\ detector acceptance. This efficiency is found not to vary significantly across the phase space. The efficiencies of the trigger requirements, $\epsilon_{\rm trig}$, and that of the reconstruction, $\epsilon_{\rm reco}$, all with respect to the preceding step, do however have significant dependence on Dalitz-plot position. The BDT, which dominates the offline selection criteria and is designed to minimise induced efficiency variations across the Dalitz plot, behaves as expected with $\epsilon_{\rm offline}$ being approximately independent of position in phase space. The total efficiency, $\epsilon_{\rm total}$, is shown in Fig.~\ref{im_efficiencyVariation} as a function of position in both standard Dalitz plot and SDP. Smooth functions are obtained by kernel estimation~\cite{Poluektov:2014rxa} and the model obtained using the SDP is used in the analysis to avoid edge effects. Given the differences between Run~1 and Run~2 data for every element of Eq.~\eqref{eqn_effs}, separate efficiency maps are used for the two data-taking periods.

\begin{figure}[!tb]
\centering
\includegraphics[width=.49\textwidth]{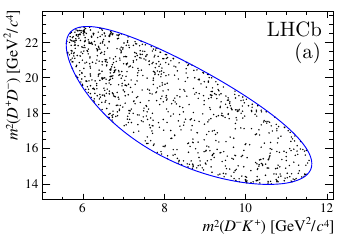}
\includegraphics[width=.49\textwidth]{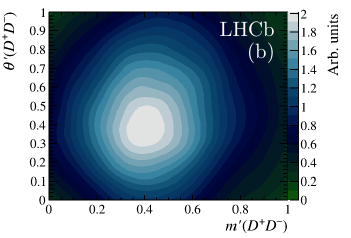}\\
\includegraphics[width=.49\textwidth]{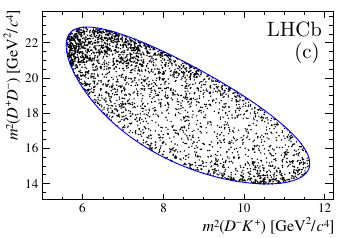}
\includegraphics[width=.49\textwidth]{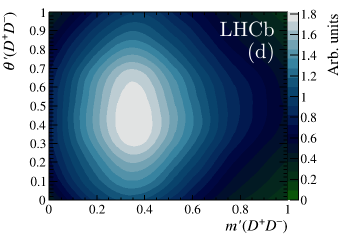}
\caption{Visualisation of the sideband candidates in the (a,c) standard Dalitz plot and (b,d)~derived background models in the SDP  for (a,b) Run~1 and (c,d) Run~2 data. \label{im_dalitzBgModels}}
\end{figure}

The residual combinatorial background contribution, though small, is accounted for in the fit. A model is derived from candidates in the high \B-candidate mass sideband, between 5.35 and 5.69\gevcc. In order to increase the sample size available for this modelling, the BDT requirement is relaxed by an amount that is seen not to influence the distribution of the background candidates in the Dalitz plot significantly. 
A kernel estimation procedure is applied to the selected background candidates to reduce the impact of statistical fluctuations. 
Due to the different selections applied to Run~1 and Run~2 data, both online and offline, separate background models are obtained for each.
The background candidates in the regular Dalitz plot are shown in Fig.~\ref{im_dalitzBgModels}, along with the derived background model as a function of SDP position, obtained using a kernel density estimation~\cite{Poluektov:2014rxa}.

\section{Amplitude model}
\label{sec_ampmodel}
\subsection{Model content}
\label{sec_ampmodel_content}

The masses of the particles involved in the \BpDpDmKp decay give rise to limits on the allowed masses of on-shell intermediate resonances: $3.74\gevcc < m\left(\Dp\Dm\right) < 4.79\gevcc$ and $2.36\gevcc< m(\Dp\Kp),m(\Dm\Kp) < 3.41\gevcc$. 
As described in Sec.~\ref{sec:Introduction}, only charmonium resonances in the $\Dp\Dm$ channel are anticipated. 
Moreover, only states with natural spin-parity ($J^P = 0^+, 1^-, 2^+, ...$) can decay strongly to a pair of pseudoscalar mesons, and resonances with very high intrinsic spin are unlikely to be produced in the decay of a pseudoscalar $\Bp$ meson. Given these considerations, the resonances initially considered are listed, with their properties, in Table~\ref{tab_ContributingResonances}.

\begin{table}[!tb]
\renewcommand{\arraystretch}{1.1}
\centering
\caption{Components which may appear in the $\Dp\Dm$ spectrum of \BpDpDmKp decays, and their properties as given by the Particle Data Group (PDG)~\cite{PDG2019}. For the $\psi(3770)$ mass and the mass/width of both the $\chi_{c2}(3930)$ and $X(3842)$, the values in Ref.~\cite{LHCb-PAPER-2019-005} are used. \label{tab_ContributingResonances}}
\resizebox{\textwidth}{!}{
\begin{tabular}{l p{4cm} r@{ $\pm$ }l r@{ $\pm$ }l}
\hline\hline
Partial wave ($J^{PC}$) & Resonance & \multicolumn{2}{l}{Mass (\nspmevcc)} & \multicolumn{2}{l}{Width (\nspmev)} \\
\hline
S~wave ($0^{++}$)	&	$\chi_{c0}(3860)$	& $3862$ & $43$	    & 	$201$ & $145$ \\
                    &   $X(3915)$           & $3918.4$ & $1.9$    &   $20$ & $5$ \\
                    \hline
P~wave ($1^{--}$)	&	$\psi(3770)$		& $3778.1$ & $0.9$	& 	$27.2$ & $1.0$ \\
					& 	$\psi(4040)$		& $4039$ & $1$		&	$80$ & $10$ \\
					& 	$\psi(4160)$		& $4191$ & $5$		&	$70$ & $10$ \\
					& 	$\psi(4260)$		& $4230$ & $8$		&	$55$ & $19$ \\
					& 	$\psi(4415)$		& $4421$ & $4$		&	$62$ & $20$ \\	
					\hline
D~wave ($2^{++}$)	&	$\chi_{c2}(3930)$	& $3921.9$ & $0.6$	& 	$36.6$ & $2.1$ \\
\hline
F~wave ($3^{--}$)   &   $X(3842)$           & $3842.71$ & $0.20$ & $2.79$ & $0.62$ \\ 
\hline\hline
\end{tabular}
}
\end{table}

Contributions to the S~wave can be expected, but there are few previous experimental results on scalar $c\bar{c}$ resonances. The Belle collaboration~\cite{Chilikin:2017evr} has reported the observation of a $\chi_{c0}(3860)$ state,\footnote{
The PDG convention, which is followed in this paper, is that the symbol used to denote a particle depends only on its quantum numbers and does not imply any interpretation of its substructure.
}
seen as a $D\Dbar$ resonance in the process $\epem \to \jpsi D\Dbar$, where the $J^{PC} = 0^{++}$ hypothesis is favoured over the $2^{++}$ hypothesis at the level of $2.5\,\sigma$. This resonance is yet to be confirmed, and there could be other states or nonresonant S-wave $D\Dbar$ contributions. The PDG listing~\cite{PDG2019} includes a $X(3915)$ state, with $J^{PC} = 0^{++}$ or $2^{++}$, seen produced in $\gamma\gamma$ collisions by the Belle~\cite{Uehara:2009tx} and BaBar~\cite{Lees:2012xs} collaborations (and also possibly in $B \to X(3915)K$ decays~\cite{Abe:2004zs,Aubert:2007vj}) and decaying to the $\jpsi\omega$ final state --- it has not been seen in the $D\Dbar$ final state. It appears that this structure may be caused by the $\chi_{c2}(3930)$ state~\cite{Zhou:2015uva}, which has also been seen by BaBar to be produced in $\gamma\gamma$ collisions~\cite{Aubert:2010ab} and has been studied more recently and precisely by LHCb in $pp$ collisions~\cite{LHCb-PAPER-2019-005}. However, the existence of both spin-0 and spin-2 states near $3930\mevcc$~\cite{Chen:2012wy,Duan:2020tsx} is not excluded. At higher mass, the $\chi_{c0}(4500)$ and $\chi_{c0}(4700)$ states have been seen as $\jpsi\phi$ resonances in an LHCb amplitude analysis of $\Bp \to \jpsi\phi\Kp$ decays~\cite{LHCb-PAPER-2016-018,LHCb-PAPER-2016-019}, with masses and widths $M =  4506\,^{+16}_{-19} \mevcc$, $\Gamma = 92\pm29 \mev$ and $M = 4704\,^{+17}_{-26} \mevcc$, $\Gamma = 120\pm50 \mev$ respectively. Given that their quantum numbers have been measured as $J^{PC} = 0^{++}$, these could in principle be seen in \BpDpDmKp decays, but since their composition is unclear it is difficult to make any prediction as to whether this is likely or not. 

A larger number of vector $c\bar{c}$ states have been observed, since these can be produced directly in $\epem$ collisions. The $\psi(3770)$, $\psi(4040)$, $\psi(4160)$ and $\psi(4415)$ states are all well established and known to decay to $D\Dbar$, therefore all might be expected to appear in \BpDpDmKp decays. 
The $\psi(3770)$ and $\psi(4160)$ resonances were included in the previous BaBar~\cite{Lees:2014abp} and Belle~\cite{Brodzicka:2007aa} amplitude analyses of the \BpDzDzbKp decay, while $\psi(4040)$ and $\psi(4415)$ components were additionally included in an LHCb amplitude analysis of $\Bp \to \Kp\mumu$ decays~\cite{LHCb-PAPER-2016-045} but found not to contribute significantly. The $\psi(4260)$ state, originally called $Y(4260)$, was observed by the BaBar collaboration through radiative return in \epem\ production to the $\jpsi\pip\pim$ final state~\cite{Aubert:2005rm}. Subsequently confirmed by CLEO, Belle and BESIII collaborations~\cite{Coan:2006rv,Yuan:2007sj,Ablikim:2016qzw}, including through direct \epem\ production, it has not been observed in the $D\Dbar$ final state, nor is there convincing evidence for its production in $B$ decays.  The only $\psi(4260)$ decays to be observed to date contain a \jpsi\ meson in the final state, although a $\psi(4230)$ state with similar mass and width ($M = 4218\,^{+\,5}_{-\,4} \mevcc$, $\Gamma = 59\,^{+\,12}_{-\,10} \mev$) has been seen by BESIII to be produced in \epem\ collisions in the $\chi_{c0}\omega$, $h_c\pip\pim$ and $\psi(2S)\pip\pim$ final states~\cite{BESIII:2016adj,Ablikim:2017oaf,Ablikim:2019apl}.  
It is sufficient to consider one of the two as a candidate contribution to the \BpDpDmKp Dalitz plot; the $\psi(4260)$ is used as it is considered to be better established in the PDG 2019 listings.\footnote{
    In its 2020 edition, the PDG has changed its treatment of the $\psi(4230)$ and $\psi(4260)$ states, but this does not impact significantly on the analysis.
} 
Two further vector states, the $\psi(4360)$ and $\psi(4660)$, have been seen in radiative return from \epem\ collisions to the $\psi(2S)\pip\pim$ final state by the BaBar and Belle collaborations~\cite{Lees:2012pv,Wang:2014hta}.
Moreover, a BESIII scan of the energy dependence of the \mbox{$\epem \to \jpsi\pip\pim$} cross-section~\cite{Ablikim:2016qzw} suggests that the structure around $4260 \mevcc$ is composed of two states: one with $M=4222.0\pm3.1\pm1.4 \mevcc, \Gamma = 44.1\pm4.3\pm2.0 \mev$ and another with $M=4320.0\pm10.4\pm7.0 \mevcc, \Gamma = 101.4\,^{+25.3}_{-19.7}\pm10.2 \mev$. In the PDG 2019 edition, the results for the first are included in the averages of the properties of the $\psi(4260)$, while those for the second are included in the $\psi(4360)$ averages.  
Both the $\psi(4360)$ and $\psi(4660)$ are considered unlikely to be present in \BpDpDmKp decays since they have never previously been observed to either be produced in $B$ decays or to decay to $D\Dbar$ final states.
They are therefore not included in Table~\ref{tab_ContributingResonances}.

In the $D$~wave, the $\chi_{c2}(3930)$ state has recently been studied by LHCb in $pp$ collisions~\cite{LHCb-PAPER-2019-005}, leading to significant improvement in the knowledge of its properties.
However, its quantum numbers are assumed, and while previous analyses have indicated a preference for a spin-2 particle in this mass range~\cite{Uehara:2005qd,Aubert:2010ab} it is not experimentally excluded that the measured structure is spin-0 or, at least, has a spin-0 contribution. Therefore, it is important to determine the spin of the $\chi_{c2}(3930)$ resonance in this analysis.

Finally, a candidate for the spin-3 $\psi_3(1^3D_3)$ charmonium state, the $X(3842)$, has recently been observed by LHCb decaying to $D\Dbar$~\cite{LHCb-PAPER-2019-005}. Its quantum numbers have not been measured, but its properties fit the expectation for that state. Production of \mbox{spin-3} states in $\B$-meson decays is suppressed, especially when there is little phase space available, and therefore this state is not expected to contribute at a significant level in \BpDpDmKp decays.

\subsection{Model development}
\begin{figure}
\centering
\includegraphics[width=.49\textwidth]{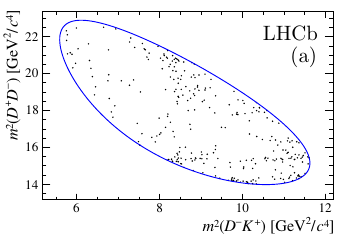}
\includegraphics[width=.49\textwidth]{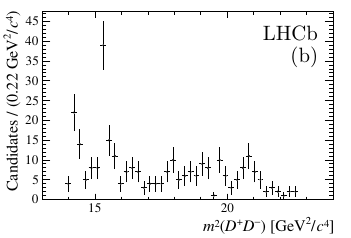}\\
\includegraphics[width=.49\textwidth]{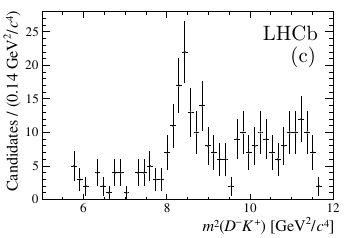}
\includegraphics[width=.49\textwidth]{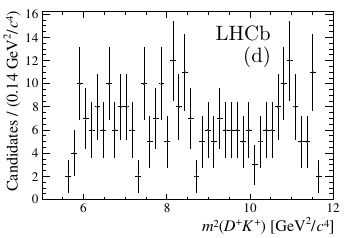}
\caption{Run~1 data entering the amplitude fit, shown in the Dalitz plot and its projection onto the invariant-mass squared for each of the three pairs of the final-state particles.}
\label{fig_RawDataDalitz_Run1}
\end{figure}

\begin{figure}
    \centering
\includegraphics[width=.49\textwidth]{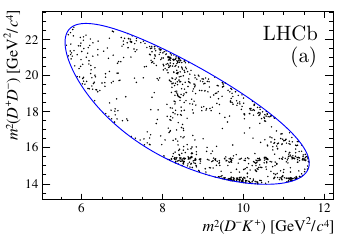}
\includegraphics[width=.49\textwidth]{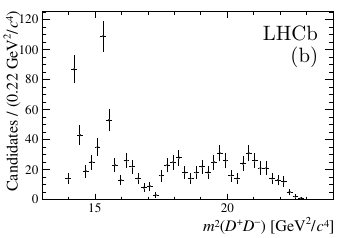}\\
\includegraphics[width=.49\textwidth]{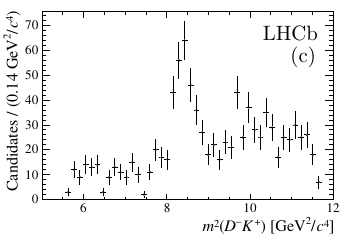}
\includegraphics[width=.49\textwidth]{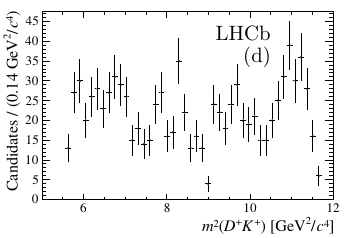}
    \caption{Run~2 data entering the amplitude fit, shown in the Dalitz plot and its projection onto the invariant-mass squared for each of the three pairs of the final-state particles.}
    \label{fig_RawDataDalitz_Run2}
\end{figure}

Selected signal candidates entering the invariant-mass fit shown in Fig.~\ref{im_massfit_and_efficiencies} are further filtered by applying a window of width $40\mevcc$ around the known \Bp mass. The 2011 and 2012 data are combined into a single Run~1 dataset, and the 2015--2018 data are combined into a single Run~2 dataset. The Dalitz plot and its projections are shown in Figs.~\ref{fig_RawDataDalitz_Run1} and~\ref{fig_RawDataDalitz_Run2}, for Run~1 and Run~2 respectively. The Dalitz-plot coordinates are determined after refitting the candidate decays, imposing the constraints that the reconstructed \Bp and \Dpm masses should match their known values and that the reconstructed \Bp meson should originate at its associated primary vertex. This improves the resolution of the Dalitz-plot coordinates; for example, the $m(\Dp\Dm)$ resolution is reduced from $10$--$13\mevcc$ to $1.5$--$3.5\mevcc$, depending upon position in the Dalitz plot.
As the resolution is much smaller than the width of the narrowest resonance considered in the analysis, it is neglected in the amplitude fit.
A simultaneous fit of the Run~1 and Run~2 datasets is carried out with separate efficiency maps, background models, and fixed signal yields for the two samples. All other model parameters are shared.

Models which reproduce the Dalitz plot distribution of the data are developed by considering resonances listed in Table~\ref{tab_ContributingResonances} and additional resonant and nonresonant components.  
The $\psi(3770) \to \Dp\Dm$ and $\chi_{c2}(3930)\to\Dp\Dm$ resonances, which are both clearly seen in the data, are taken as a starting point.
Further components are included in the model if they cause a significant reduction in the negative log-likelihood obtained from the fit to data, while not causing instabilities in the fit or producing excessively large inference effects and hence a sum of fit fractions far from 100\%.
The complex coefficients associated with all resonant or nonresonant components are allowed to vary freely, with the exception of that for the $\psi(3770) \to \Dp\Dm$ component, which is fixed to unit length along the real axis to serve as a reference amplitude.
The masses and widths of contributing resonances are all allowed to vary, though Gaussian constraints, with parameters corresponding to the central values and uncertainties in Table~\ref{tab_ContributingResonances}, are applied to those of the $\psi(3770)$, $\psi(4040)$, $\psi(4160)$, and $\psi(4415)$ states.

It is observed that significantly better agreement between the model and the data is obtained when including a spin-0 $D\Dbar$ component that overlaps with the $\chi_{c2}(3930)$ state, labelled $\chi_{c0}(3930)$.
The presence of a spin-0 component in this $\chi_{cJ}(3930)$ region may mean that previous measurements of the mass and width of the $\chi_{c2}(3930)$ state, based on an assumption of a single resonance, are not reliable.
Therefore, the masses and widths of both the spin-0 and spin-2 components are allowed to vary freely.

It is found that the inclusion of at least one nonresonant component is essential to obtain a good fit to data. 
A number of parameterisations are considered, including the case of completely uniform Dalitz plot density and modulation of the nonresonant amplitude by either polynomial or exponential form factors, and the possibility of a spin-1, instead of spin-0, angular term.
A quasi-model-independent partial wave description of the S~wave, as used for example in Refs.~\cite{Aitala:2005yh,LHCb-PAPER-2016-026,LHCb-PAPER-2019-017}, is also attempted, but is not viable with the current sample sizes.  
In all cases, parameters associated with the nonresonant model are allowed to vary freely in the fit to data.

For each configuration, the minimisation is repeated 100 times, randomising the starting parameters at each iteration. The minimisation that is consistently found to yield the best likelihood value is selected.
In order to assess the fit quality, a $\chi^2$ computation is performed, with an adaptive binning scheme ensuring a minimum of 20 candidates in each bin. The associated number of degrees of freedom is determined using an ensemble of pseudoexperiments generated at the fit minimum. The goodness of fit is assessed using this figure of merit as well as the change in negative log-likelihood value between different configurations.

\section{Results}
\label{sec:results}

\subsection{\boldmath Model excluding $\Dm\Kp$ resonances}
\label{sec:results_noDK}

The data in Figs.~\ref{fig_RawDataDalitz_Run1} and~\ref{fig_RawDataDalitz_Run2} exhibit a striking excess at $m^2(\Dm\Kp)\approx 8.25\gevgevcccc$, in both Run~1 and Run~2, which cannot be accounted for by introducing resonances only in the $\Dp\Dm$ decay channel. To illustrate this, the first model presented excludes any resonant content from the $\Dm\Kp$ channel. The model includes the $\psi(3770)$, $\chi_{c0}(3930)$, $\chi_{c2}(3930)$, $\psi(4040)$, $\psi(4160)$, and $\psi(4415)$ resonances, which are necessary to describe structure in the $m(\Dp\Dm)$ spectrum. 
A nonresonant component is included and described by an exponential S-wave lineshape in the $\Dm\Kp$ spectrum. 

\begin{figure}[!tb]
\centering
\includegraphics[width=.49\textwidth]{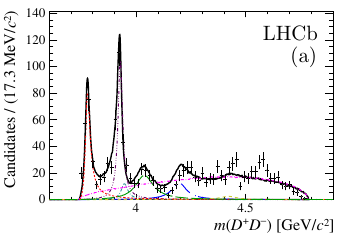}
\includegraphics[width=.49\textwidth]{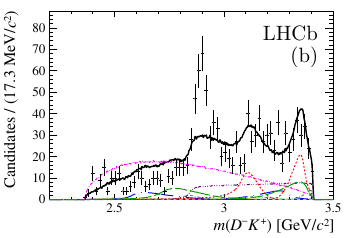}\\
\includegraphics[width=.49\textwidth]{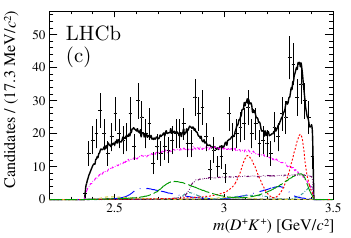}
\includegraphics[width=.49\textwidth]{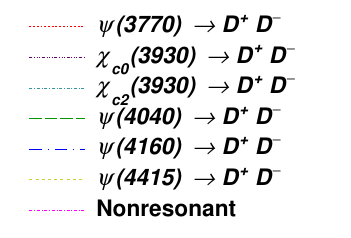}
\caption{Comparisons of the invariant-mass distributions of \BpDpDmKp candidates to the fit projections without any resonant component in the $\Dm\Kp$ channel.
The total fit function (solid black line) and contributions from individual components (non-solid coloured lines) are shown as detailed in the legend. \label{im_model1_noDK_BmDpDmKm}}
\end{figure}

\begin{figure}[!tb]
\centering
\includegraphics[width=.8\textwidth]{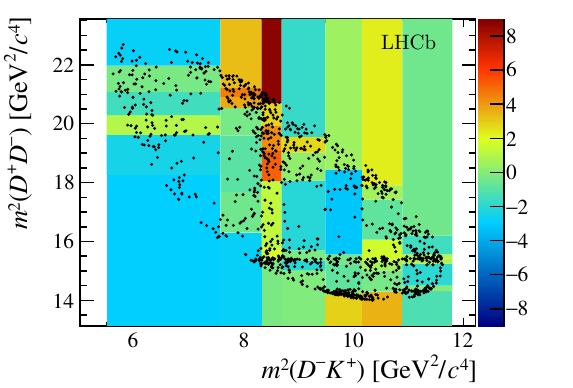}
\caption{Normalised residual between the data and the model excluding any $\Dm\Kp$ components, shown across the Dalitz plot with a minimum of 20 data entries in each bin. 
\label{im_model1_noDK_BmDpDmKm_DalitzGoF}}
\end{figure}

\begin{figure}[!tb]
\centering
\includegraphics[width=.8\textwidth]{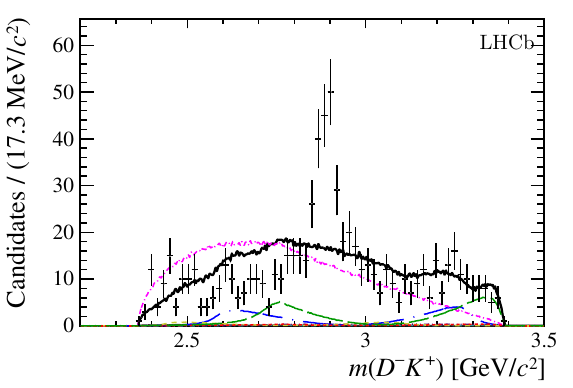}
\caption{Comparison of the $m(\Dm\Kp)$ distribution and the fit projection for a model excluding any $\Dm\Kp$ resonance, after requiring $m(\Dp\Dm)>4\gevcc$ to suppress reflections from charmonium resonances.    The different components are shown as indicated in the legend of Fig.~\ref{im_model1_noDK_BmDpDmKm}. \label{im_model1_noDK_BmDpDmKm_restricted}}
\end{figure}

The Dalitz-plot projections from this fit are compared to the data in Fig.~\ref{im_model1_noDK_BmDpDmKm}.  Contributions from individual components are superimposed. The goodness of fit is quantified in Fig.~\ref{im_model1_noDK_BmDpDmKm_DalitzGoF}, where the largest deviations are seen in the $m^2(\Dm\Kp)\approx 8.25\gevgevcccc$ region of the Dalitz plot. To illustrate this more clearly, a comparison between data and the result of the fit is made in Fig.~\ref{im_model1_noDK_BmDpDmKm_restricted} after excluding low-mass charmonium resonances through the requirement $m(\Dp\Dm)>4\gevcc$. 

It is concluded that a satisfactory description of the data cannot be obtained without including one or more components that model structure in $m(\Dm\Kp)$ explicitly.
The same conclusion is reached with a model-independent analysis, as described in Ref.~\cite{LHCb-PAPER-2020-024}.

\newcommand{\noDKFitTag}{20-05May-15_Variant2_1_2_3_4_5_27_29_lasso_-1_00000_bwradconf_0}
\newcommand{\nominalFitTag}{20-05May-08_Variant20_constrPsis_1_2_3_4_5_13_27_29_12_lasso_-1_00000}

\subsection{\boldmath Baseline model including $\Dm\Kp$ resonances}

The simplest way to account for the $m(\Dm\Kp)$ structure is by adding resonances to the model.
Analysis of the current data sample cannot, however, exclude the possibility that hadronic effects such as rescattering may be important, in particular given the observation that the structure appears near to the $D^*K^*$ threshold.
More detailed investigations of plausible explanations for the observed structure will require new theoretical models to be developed and larger data samples to be analysed.

The baseline model includes the same components as in Sec.~\ref{sec:results_noDK}, but adds both spin-1 and spin-0 $\Dm\Kp$ resonances. An exponential S-wave lineshape in the $\Dm\Kp$ channel remains the best description of the nonresonant contribution. The projections of the Dalitz plot, with fit results superimposed, are shown in Fig.~\ref{im_model1_BmDpDmKm}. In Appendix~\ref{app:helicity}, the results are compared to the helicity-angle distributions in eight bins of the invariant-mass distribution of each pair of particles. A comparison to the distributions of the angular moments (defined in Ref.~\cite{LHCb-PAPER-2020-024}) of each pair of particles is made in Appendix~\ref{app:moments}. The results for the fit parameters and the fit fractions for each component are shown in Tables~\ref{tab_baselineModelSummary} and~\ref{tab_baselineModelSummary_shapeParams}, where $X_1(2900)$ and $X_0(2900)$ are used to label the new spin-1 and spin-0 $\Dm\Kp$ states, respectively. 
These results include systematic uncertainties, the evaluation of which is described in Sec.~\ref{sec_systematics}.
The coefficient of the nonresonant exponential lineshape is found to be $(\round{2}{0.0809} \pm \round{2}{0.0496})~(\nspgevgevcccc)^{-1}$, where the uncertainty is statistical only. 
The interference fit fractions are given in Table~\ref{tab_baselineModelSummary_intFracs}, with their statistical and systematic uncertainties.

\begin{figure}[!tb]
\centering
\includegraphics[width=.49\textwidth]{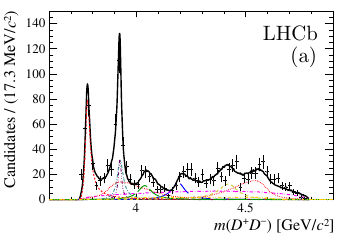}
\includegraphics[width=.49\textwidth]{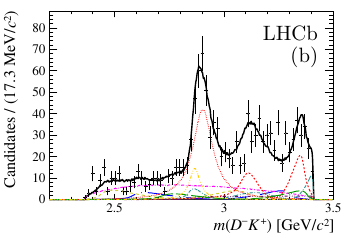}\\
\includegraphics[width=.49\textwidth]{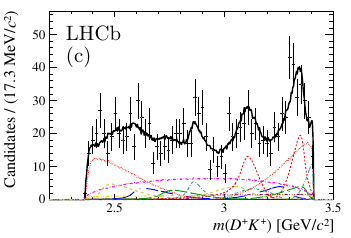}
\includegraphics[width=.49\textwidth]{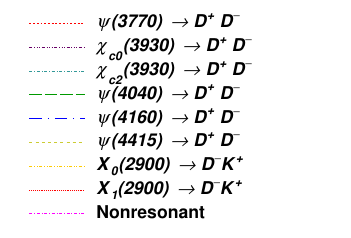}
\caption{
Comparisons of the invariant-mass distributions of \BpDpDmKp candidates in data to the fit projection of the baseline model.
The total fit function and contributions from individual components are shown as detailed in the legend. \label{im_model1_BmDpDmKm}}
\end{figure}

\begin{table}[!tb]
\renewcommand{\arraystretch}{1.1}
\centering
\caption{Magnitude and phase of the complex coefficients in the amplitude model, together with fit fractions for each component. The quantities are reported after correction for fit biases (see Sec.~\ref{sec_systematics}). The first uncertainty is statistical and the second is the sum in quadrature of all systematic uncertainties.
\label{tab_baselineModelSummary}}
\begin{tabular}{l r@{ $\pm$ }l@{ $\pm$ }l r@{ $\pm$ }l@{ $\pm$ }l r@{ $\pm$ }l@{ $\pm$ }l}
\hline\hline
Resonance & \multicolumn{3}{c}{Magnitude} & \multicolumn{3}{c}{Phase (rad)} & \multicolumn{3}{c}{Fit fraction (\%)}\\
\hline
\multicolumn{2}{l}{$\Dp\Dm$ resonances}\\
$\psi(3770)$ & \multicolumn{3}{c}{1 (fixed)} & \multicolumn{3}{c}{0 (fixed)} & $14.5$ & $1.2$ & $0.8$\\
$\chi_{c0}(3930)$ & $\round{2}{0.506}$ & $\round{2}{0.064}$ & $\round{2}{0.017}$ & $\round{2}{2.162}$ & $\round{2}{0.184}$ & $\round{2}{0.034}$ & $3.7$ & $0.9$ & $0.2$\\
$\chi_{c2}(3930)$ & $\round{2}{0.703}$ & $\round{2}{0.064}$ & $\round{2}{0.012}$ & $\round{2}{0.827}$ & $\round{2}{0.170}$ & $\round{2}{0.133}$ & $7.2$ & $1.2$ & $0.3$\\
$\psi(4040)$ & $\round{2}{0.585}$ & $\round{2}{0.078}$ & $\round{2}{0.035}$ & $\round{2}{1.416}$ & $\round{2}{0.176}$ & $\round{2}{0.084}$ & $5.0$ & $1.3$ & $0.4$\\
$\psi(4160)$ & $\round{2}{0.668}$ & $\round{2}{0.084}$ & $\round{2}{0.052}$ & $\round{2}{0.898}$ & $\round{2}{0.225}$ & $\round{2}{0.092}$ & $6.6$ & $1.5$ & $1.2$\\
$\psi(4415)$ & $\round{2}{0.797}$ & $\round{2}{0.080}$ & $\round{2}{0.061}$ & $\round{2}{-1.458}$ & $\round{2}{0.197}$ & $\round{2}{0.091}$ & $9.2$ & $1.4$ & $1.5$\\
\hline\multicolumn{2}{l}{$\Dm\Kp$ resonances}\\
$X_0(2900)$ & $\round{2}{0.619}$ & $\round{2}{0.079}$ & $\round{2}{0.025}$ & $\round{2}{1.091}$ & $\round{2}{0.193}$ & $\round{2}{0.095}$ & $5.6$ & $1.4$ & $0.5$\\
$X_1(2900)$ & $\round{2}{1.449}$ & $\round{2}{0.086}$ & $\round{2}{0.032}$ & $\round{2}{0.367}$ & $\round{2}{0.102}$ & $\round{2}{0.049}$ & $30.6$ & $2.4$ & $2.1$\\
\hline
Nonresonant & $\round{2}{1.293}$ & $\round{2}{0.088}$ & $\round{2}{0.043}$ & $\round{2}{-2.410}$ & $\round{2}{0.119}$ & $\round{2}{0.508}$ & $24.2$ & $2.2$ & $0.5$\\
\hline\hline 
\end{tabular}
\end{table}

\begin{table}[!tb]
\renewcommand{\arraystretch}{1.1}
\centering
\caption{Lineshape parameters for the $\chi_{c0,2}(3930)$ and $X_{0,1}(2900)$ resonances determined from the fit. The first uncertainty is statistical and the second is the sum in quadrature of all systematic uncertainties.
\label{tab_baselineModelSummary_shapeParams}}
\begin{tabular}{l r@{ $\pm$ }l@{ $\pm$ }l r@{ $\pm$ }l@{ $\pm$ }l }
\hline\hline
Resonance & \multicolumn{3}{c}{Mass (\nspgevcc)} & \multicolumn{3}{c}{Width (\nspmev)} \\
\hline
$\chi_{c0}(3930)$ & $3.9238$ & $0.0015$ & $0.0004$ & $17.4$ & $\phantom{1}5.1$ & $0.8$\\
$\chi_{c2}(3930)$ & $3.9268$ & $0.0024$ & $0.0008$ & $34.2$ & $\phantom{1}6.6$ & $1.1$\\
\hline
$X_0(2900)$ & $\round{3}{2.8663}$ & $\round{3}{0.0065}$ & $\round{3}{0.0020}$ & $\round{0}{57.2}$ & $\round{0}{12.2}$ & $\round{0}{4.1}$ \\
$X_1(2900)$ & $\round{3}{2.9041}$ & $\round{3}{0.0048}$ & $\round{3}{0.0013}$ & $\round{0}{110.3}$ & $\round{0}{10.7}$ & $\round{0}{4.3}$ \\
\hline\hline
\end{tabular}
\end{table}

\begin{sidewaystable}
\caption{Interference fit fractions (\%) obtained from the results of the amplitude fit with the baseline model. Uncertainties are statistical and systematic, respectively. Absent entries correspond to pairs of resonances that do not interfere, because they either inhabit separate regions of phase space or belong to different partial waves in the same two-body combination. \label{tab_baselineModelSummary_intFracs}}
\resizebox{\textwidth}{!}{
\begin{tabular}{l  r@{ $\pm$ }l@{ $\pm$ }l  r@{ $\pm$ }l@{ $\pm$ }l  r@{ $\pm$ }l@{ $\pm$ }l  r@{ $\pm$ }l@{ $\pm$ }l  r@{ $\pm$ }l@{ $\pm$ }l  r@{ $\pm$ }l@{ $\pm$ }l  r@{ $\pm$ }l@{ $\pm$ }l  r@{ $\pm$ }l@{ $\pm$ }l  r@{ $\pm$ }l@{ $\pm$ }l  r@{ $\pm$ }l@{ $\pm$ }l }
\hline\hline
 & \multicolumn{3}{c}{$\chi_{c0}(3930)$} & \multicolumn{3}{c}{$\chi_{c2}(3930)$} & \multicolumn{3}{c}{$\psi(4040)$} & \multicolumn{3}{c}{$\psi(4160)$} & \multicolumn{3}{c}{$\psi(4415)$} & \multicolumn{3}{c}{$X_0(2900)$} & \multicolumn{3}{c}{$X_1(2900)$} & \multicolumn{3}{c}{Nonresonant}\\
\hline
$\psi(3770)$ & \multicolumn{3}{c}{---}  & \multicolumn{3}{c}{---}  & \phm$3.7$ & $0.5$ & $0.1831$ & \phm$1.7$ & $0.6$ & $0.3$ & $-3.3$ & $0.7$ & $0.6$ & $-0.6$ & $0.2$ & $0.1$ & $-4.6$ & $0.5$ & $0.6$ & $-0.4$ & $0.3$ & $0.5$\\
$\chi_{c0}(3930)$ & \multicolumn{3}{l}{}  & \multicolumn{3}{c}{---}  & \multicolumn{3}{c}{---}  & \multicolumn{3}{c}{---}  & \multicolumn{3}{c}{---}  & \phm$0.1$ & $0.1$ & $0.0$ & \phm$0.5$ & $0.1$ & $0.0$ & \phm$3.2$ & $0.7$ & $1.5$\\
$\chi_{c2}(3930)$ & \multicolumn{3}{l}{}  & \multicolumn{3}{l}{}  & \multicolumn{3}{c}{---}  & \multicolumn{3}{c}{---}  & \multicolumn{3}{c}{---}  & $-0.2$ & $0.0$ & $0.1$ & $-1.5$ & $0.2$ & $0.4$ & \phm$0.0$ & $0.0$ & $0.0$\\
$\psi(4040)$ & \multicolumn{3}{l}{}  & \multicolumn{3}{l}{}  & \multicolumn{3}{l}{}  & $-1.2$ & $1.3$ & $0.1$ & $-0.3$ & $0.7$ & $0.3$ & \phm$0.1$ & $0.1$ & $0.0$ & $-0.6$ & $0.5$ & $0.2$ & $-0.6$ & $0.4$ & $0.5$\\
$\psi(4160)$ & \multicolumn{3}{l}{}  & \multicolumn{3}{l}{}  & \multicolumn{3}{l}{}  & \multicolumn{3}{l}{}  & $-5.1$ & $1.3$ & $0.9$ & $-0.2$ & $0.1$ & $0.0$ & $-2.8$ & $0.5$ & $0.4$ & $-0.2$ & $0.2$ & $0.4$\\
$\psi(4415)$ & \multicolumn{3}{l}{}  & \multicolumn{3}{l}{}  & \multicolumn{3}{l}{}  & \multicolumn{3}{l}{}  & \multicolumn{3}{l}{}  & \phm$0.0$ & $0.1$ & $0.1$ & \phm$3.1$ & $0.5$ & $0.2$ & \phm$0.5$ & $0.3$ & $0.4$\\
$X_0(2900)$ & \multicolumn{3}{l}{}  & \multicolumn{3}{l}{}  & \multicolumn{3}{l}{}  & \multicolumn{3}{l}{}  & \multicolumn{3}{l}{}  & \multicolumn{3}{l}{}  & \multicolumn{3}{c}{---}  & \phm$2.3$ & $1.4$ & $3.0$\\
$X_1(2900)$ & \multicolumn{3}{l}{}  & \multicolumn{3}{l}{}  & \multicolumn{3}{l}{}  & \multicolumn{3}{l}{}  & \multicolumn{3}{l}{}  & \multicolumn{3}{l}{}  & \multicolumn{3}{l}{}  & \multicolumn{3}{c}{---} \\
\hline\hline
\end{tabular}
}
\end{sidewaystable}

\begin{figure}[!tb]
\centering
\includegraphics[width=.49\textwidth]{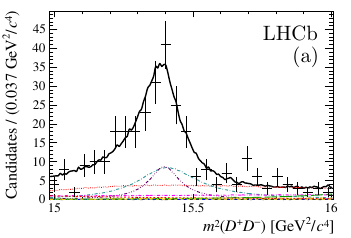}
\includegraphics[width=.49\textwidth]{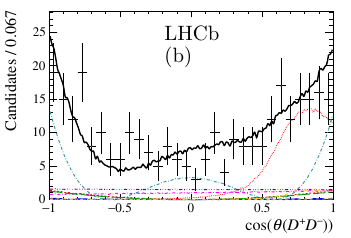}
\caption{Comparison of the data and fit projection in the $\chi_{cJ}(3930)$ region, shown for the (a)~$\Dp\Dm$ invariant-mass squared and (b) helicity angle.  The different components are shown as indicated in the legend of Fig.~\ref{im_model1_BmDpDmKm}. \label{im_model1_BmDpDmKm_chic2region}}
\end{figure}

As described in Sec.~\ref{sec_ampmodel_content}, $D\Dbar$ resonant structure has previously been observed in the $\chi_{cJ}(3930)$ region, however it has usually been assumed to arise from the $\chi_{c2}(3930)$
resonance.
The mass and helicity-angle distributions of candidates in this region, shown in Fig.~\ref{im_model1_BmDpDmKm_chic2region}, clearly demonstrate that both spin-0 and spin-2 contributions are necessary.
The masses and widths of these two components are completely free to vary in the fit; they are found to have consistent masses while the fit prefers a narrower width for the spin-0 state.
If both spin-0 and spin-2 states are present at the same mass, one would generically expect the spin-0 state to be broader since its decay to a $\Dp\Dm$ pair is in S~wave, as compared to D~wave for the spin-2 state, and therefore is not suppressed by any angular momentum barrier.
This expected pattern is seen in some explicit calculations of the properties of the $\chi_{cJ}(2P)$ states~\cite{Guo:2012tv}, however the observed pattern is consistent with other theoretical predictions~\cite{Duan:2020tsx}.
Moreover, the fitted $\chi_{c0}(3930)$ parameters are consistent with those of the $X(3915)$ state.

The $\chi_{c0}(3930)$ state is the only component in the $\Dp\Dm$ S~wave in the baseline model.
The broad $\chi_{c0}(3860)$ state, reported by the Belle collaboration~\cite{Chilikin:2017evr}, has been included in alternative fit models but is disfavoured. Fits in which additional S-wave structure is introduced \eg\ through a nonresonant component, have been attempted but tend to destabilise the fit, which is understood as a consequence of there being too much freedom in the S~wave.
In fact the nonresonant component in the $\Dm\Kp$ projection covers most of the $m(\Dp\Dm)$ range, as can be seen in Fig.~\ref{im_model1_BmDpDmKm} top row, but only allows a small contribution at low $m(\Dp\Dm)$ values.

A good description of the intermediate $m(\Dp\Dm)$ region is obtained by including the $\psi(4040)$, $\psi(4160)$, and $\psi(4415)$ contributions, together with reflections from the $\Dm\Kp$ structures.  Inclusion of the $\psi(4260)$ resonance was also considered during the model-building process, but its inclusion together with the $\psi(4160)$ state leads to fit instabilities, due to the similarity of their masses and widths. Between the two, a slight preference was visible in negative log-likelihood value for the $\psi(4160)$ component.

\begin{figure}[!tb]
\centering
\includegraphics[width=.49\textwidth]{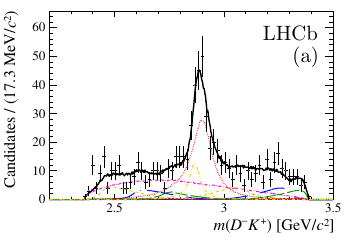}
\includegraphics[width=.49\textwidth]{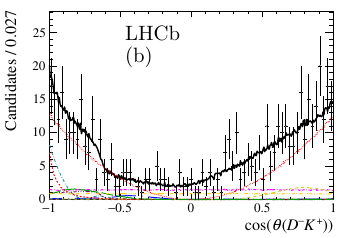}
\caption{
Comparison of data and the fit projection of the baseline model, for (a) the $\Dm\Kp$ invariant-mass distribution requiring $m(\Dp\Dm)>4\gevcc$ to suppress reflections from charmonium resonances and (b) helicity angle in the region $2.75\gevcc<m(\Dm\Kp)<3.05\gevcc$. The different components are shown as indicated in the legend of Fig.~\ref{im_model1_BmDpDmKm}. \label{im_model1_BmDpDmKm_restricted}}
\end{figure}

The impact of the $X_1(2900)$ and $X_0(2900)$ states on the agreement between the data and the model is highlighted in Fig.~\ref{im_model1_BmDpDmKm_restricted}(a) by restricting the phase space to exclude low mass charmonium resonances in the same way as in Fig.~\ref{im_model1_noDK_BmDpDmKm_restricted}.
The need for both spin-1 and spin-0 components is seen in the helicity-angle distribution shown in Fig.~\ref{im_model1_BmDpDmKm_restricted}(b).

\subsection{Other models}
\label{sec:other_models}

Numerous variations in the composition of the decay amplitude are considered in the process of establishing the baseline model. 
These include consideration of one or two states with different spins in the $\chi_{cJ}(3930)$ region, and zero, one or two states in the $X(2900)$ region, as well as the inclusion of a contribution from the $X(3842)$ state (assumed to be spin 3).
The impact of these different model choices on the negative log-likelihood resulting from the fit is summarised in Table~\ref{tab_othermodels}.
Models with two components with the same spin in the same two-body combination, and with freely varying masses and widths, tend to make the fit unstable and are therefore not included.
Similarly, variations in the description of the nonresonant component that destabilise the fit are not included as the obtained negative log-likelihood values are not reliable.

\begin{table}[!tb]
\centering
\renewcommand{\arraystretch}{1.1}
\caption{Model variations and the associated negative log-likelihood (NLL) and $\chi^2$ values.\label{tab_othermodels}}
\begin{tabular}{l c r}
\hline\hline
Model & NLL & $\chi^2$ \\
\hline
Baseline & $-3540$ & 86.1\\
\hline
Variations to $\chi_{cJ}(3930)$ region\\
$\chi_{c0}(3930)$ only  & $-$3508 & 104.2\\
$\chi_{c2}(3930)$ only  & $-$3502 & 111.1\\
$\chi_{c0}(3930)$ + $\psi(3930)$ & $-$3540 & 94.0 \\
\hline
Variations in $\Dm\Kp$ channel\\
No $\Dm\Kp$ resonances  & $-$3382 & 288.9\\
One $\Dm\Kp$ resonance (spin-0) & $-$3491 & 175.8\\
One $\Dm\Kp$ resonance (spin-1) & $-$3497 & 107.2\\
One $\Dm\Kp$ resonance (spin-2) & $-$3463 & 152.6\\
Two $\Dm\Kp$ resonances (spin-1 + spin-2) & $-3536$ & 91.6\\
\hline
Other \\
Addition of $X(3842)$ & $-$3541 & 85.3\\
\hline\hline
\end{tabular}
\end{table}

Among the models with variations to the description of the $\chi_{cJ}(3930)$ region, those including a spin-1 state (denoted $\psi(3930)$) are considered unlikely since any vector state in this region would have been seen by previous experiments, as discussed in Sec.~\ref{sec_ampmodel_content}.
Moreover, including such a state in the model, either by itself or together with a $\chi_{c2}(3930)$ state, has a large impact on other components of the model.
The $X_1(2900)$ component moves to higher mass and much broader width, with the nonresonant lineshape also changing significantly. 
These models are therefore excluded from Table~\ref{tab_othermodels}.
The model with  $\chi_{c0}(3930)$+$\psi(3930)$ states does not suffer this problem but, like other models including a $\psi(3930)$ component, has large interference effects due to the overlap between spin-1 states in the model.
This causes a higher sum of fit fractions compared to the baseline model.
All models containing the $\psi(3930)$ are thus disfavoured, leaving the approach of including $\chi_{c0}(3930)$ and $\chi_{c2}(3930)$ states as the only candidate to describe the data in the $\chi_{cJ}(3930)$ region.

Among the variations in the $\Dm\Kp$ channel, the need for two states is clear from the improvement in the NLL and $\chi^2$ values.  
Noting the proximity to the $D^*K^*$ threshold, a model with spin-0 and spin-2 states is theoretically well motivated.
However, when the masses and widths of the states are allowed to vary freely in the fit, the spin-2 component takes an extremely large ($>500\mev$) width, effectively becoming a nonresonant spin-2 component.  
While this may be due to residual imperfections in the model (discussed below), this configuration cannot be considered further in the current analysis and is therefore excluded from Table~\ref{tab_othermodels}.
Studies of larger data samples may help to shed light on whether it is possible to describe the structure in $m(\Dm\Kp)$ with spin-0 and spin-2 components.  
A model with spin-1 and spin-2 $\Dm\Kp$ resonances gives comparable, but less favourable, goodness-of-fit indicators to the baseline model.

The model with the inclusion of the $X(3842)$ state, assumed to be spin-3, demonstrates that there is no significant contribution from that component.
This supports the assumption, made in Ref.~\cite{LHCb-PAPER-2020-024}, that only states of spin up to 2 are present in $\Bp\to\Dp\Dm\Kp$ decays.
Fits with this model are, for simplicity, made neglecting resolution effects since this is done for all other fits.
If the narrow $X(3842)$ state were present in the data it would be necessary to account for resolution effects properly, but the fit neglecting them is sufficient to confirm qualitatively the absence of this contribution at any significant level.

\subsection{Residual imperfections in the baseline model}

\begin{figure}[!tb]
\centering
\includegraphics[width=.8\textwidth]{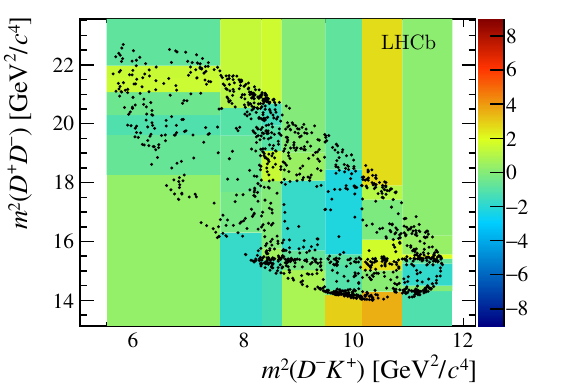}
\caption{Normalised residual between the data and the baseline model including $\Dm\Kp$ resonances, shown across the Dalitz plot with a minimum of 20 entries in each bin. 
\label{im:model1_BmDpDmKm_DalitzGoF}}
\end{figure}

\begin{figure}[!tb]
\centering
\includegraphics[width=.49\textwidth]{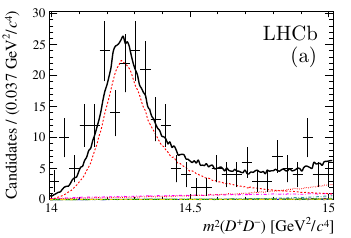}
\includegraphics[width=.49\textwidth]{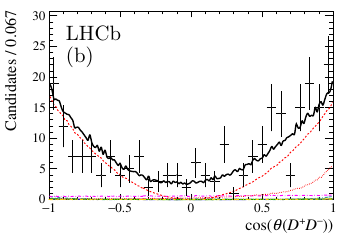}
\caption{Comparison of the data and fit projection in the region of the $\psi(3770)$ states, shown for the $\Dp\Dm$ (left) invariant-mass squared and (right) helicity angle.  The different components are shown as indicated in the legend of Fig.~\ref{im_model1_BmDpDmKm}. \label{im_model1_BmDpDmKm_psi3770region}}
\end{figure}

The goodness of fit is visualised using the binned normalised residual distribution in Fig.~\ref{im:model1_BmDpDmKm_DalitzGoF}. 
The $\chi^2$/ndf is 86.1/38.3 = 2.25, where the number of degrees of freedom, ndf, is an effective value obtained from pseudoexperiments and only statistical uncertainties are considered.
While an overall reasonable description of the data is achieved with the baseline model, there are regions of the Dalitz plot where significant imperfections remain.
The largest contributions to the binned $\chisq$ are at $\left(m^2(\Dm\Kp),m^2(\Dp\Dm)\right) \sim (10.5 \gevgevcccc, 13.5 \gevgevcccc)$ and $\sim (10.5 \gevgevcccc, 18.5 \gevgevcccc)$.
The disagreement in the first of these regions can also be seen in the \Dp\Dm helicity angle distribution at low $m^2(\Dp\Dm)$, shown in Fig.~\ref{im_model1_BmDpDmKm_psi3770region}, which shows a clear asymmetry most likely originating from interference between the $\psi(3770)$ P-wave state and S-wave $\Dp\Dm$ structure.
Since the baseline model has only very limited S~wave in this region, the asymmetry observed in data cannot be reproduced in the model.
This disagreement can also be seen in some other projections, for example at high $m(\Dm\Kp)$ in the projection of the whole Dalitz plot (Fig.~\ref{im_model1_BmDpDmKm}).

The second of the aforementioned regions of data-model disagreement corresponds to low values of $m^2(\Dp\Kp)$.
No particular disagreement is seen in other projections of this region, and therefore it is not considered a source of concern.
There does seem to be some disagreement at high $m(\Dp\Dm)$ values (Fig.~\ref{im_model1_BmDpDmKm}), but this does not make a large contribution to the $\chisq$ value.  
While the region around the $\psi(4415)$ resonance does not appear to be perfectly modelled in the projection, it is probable that at least some of this is statistical, since a very sharp structure at $m(\Dp\Dm) \sim 4.47 \gevcc$ seems unlikely to be physical. 

In summary, while the baseline model does not perfectly reproduce the observed Dalitz-plot distribution, it gives the best description of the currently available data, with a stable fit, among a large range of considered models.
Analysis of a larger sample in future will be of great interest to resolve issues associated with the imperfections of the baseline model, as will improved knowledge of $\Dp\Dm$ and $\Dm\Kp$ structures that may be obtained by analysis of other systems.

\section{Systematic uncertainties}
\label{sec_systematics}
Systematic uncertainties arising from a variety of sources are investigated, and their impact on the model amplitudes, phases, and fit fractions is quantified.
The effects on the masses and widths of resonances that are determined from the fit are also evaluated. 
Sources of systematic uncertainty are separated into those related to experimental effects and those related to model composition.
The various systematic uncertainties on the complex coefficients and fit fractions are detailed in Table~\ref{tab:syst_summary}, while those on the masses and widths of resonances are given in Table~\ref{tab:syst_shapeParams_summary}.

The yield of the signal component in the amplitude fit is fixed according to the results of the invariant-mass fit. Repeats of the amplitude fit to data are performed where the signal yield is varied, each time being sampled from a Gaussian PDF centred at the value obtained from the invariant-mass fit, having a width equal to the statistical uncertainty on that yield. The RMS of the values of the fit parameters is taken as the systematic uncertainty. The magnitude of this uncertainty is negligible, and it is therefore omitted from Table~\ref{tab:syst_summary}.

The PDF used to model the signal component in the invariant-mass fit may be imperfect. A conservative estimate of the impact of mismodelling the signal shape is obtained by replacing the DSCB shape by a simple Gaussian function. The deviation of the fit parameters from their nominal values is taken as an estimate of the systematic uncertainty.

The size of the sideband sample limits knowledge of the residual background model in the amplitude fit. 
An ensemble of bootstrapped sideband data is prepared, from which an ensemble of background models is extracted. Repeated fits to data using the different models are performed, and the RMS of the fit parameters in the resulting ensemble of fit results is taken to represent the systematic uncertainty. This uncertainty is negligible, and is therefore omitted from Table~\ref{tab:syst_summary}.

The effect of the limited size of the simulated samples, used to determine the efficiency model, is quantified. A large ensemble of simulated samples is prepared by bootstrapping the original sample, such that variations within the ensemble are representative of statistical fluctuations expected for the size of that sample. For each variant the efficiency is obtained for Run~1 and Run~2 in the same way as for the nominal efficiency model. The fit to data is then repeated once per efficiency-model-variant, and the RMS of the values of the fit parameters is taken to represent the systematic uncertainty.

The PID response in the data is obtained from calibration samples.
The systematic uncertainty incurred through this procedure principally arises from the kernel width used in the estimation of the PDFs. An alternative PID response is simulated using an alternative kernel estimation with changed width, and the efficiency models are regenerated. The fit to data is repeated with these alternative efficiency models in place and the absolute change in the fit parameters is taken as the systematic uncertainty. This uncertainty is omitted from Table~\ref{tab:syst_summary} since it is negligible.

The hardware-level trigger decision is not expected to be perfectly modelled in the simulated samples. To estimate the impact of this mismodelling of this trigger, a correction obtained from data control samples is applied to the efficiency map.
The fit is repeated with this alternative efficiency map and displacement in each parameter is computed.
This procedure overestimates the effect, since the mismodelling only affects the efficiency for candidates triggered by hardware-level hadron requirement.
Each displacement is therefore scaled according to the fraction of such candidates (64\%) to evaluate the systematic uncertainty.

The default Blatt--Weisskopf barrier radii for the parent and intermediate resonances are set to $4.0\gev^{-1}$. To evaluate the systematic uncertainty arising from the fixed radii, the fit to data is repeated where the radius for each category --- parent, charmonia or $\Dm\Kp$ resonances --- is sampled randomly from a Gaussian distribution centred at $4\gev^{-1}$ and with a width of $1\gev^{-1}$, which is the approximate size of the uncertainty on the Blatt--Weisskopf barrier radii measured in comparable systems~\cite{LHCb-PAPER-2014-036}. The RMS of the values of the fit parameters under these perturbations is taken to represent the systematic uncertainty, where the largest effect is seen when varying the Blatt--Weisskopf barrier radius of the charmonium resonances, which dominate the model.
This is the largest systematic uncertainty for several of the parameters determined from the fit.

The baseline model includes contributions that are clearly established, but the true amplitude may include components that are not significant at the current level of precision and which are consequently omitted.
In addition, the most appropriate way to model some of the components is not established, and mismodelling is a source of potential systematic uncertainty.
While many possible model variations could be considered, including too many would lead to an artificial inflation of the uncertainty.
Therefore this procedure is limited to specific variations in the partial waves where the modelling uncertainty is largest.
With reference to the discussion in Sec.~\ref{sec_ampmodel_content}, these are
\begin{itemize}
    \item $\Dp\Dm$ S~wave: Inclusion of an additional constant nonresonant component. Introducing such a component with a freely varying complex coefficient, alongside the existing nonresonant shape, destabilises the fit so instead the amplitude and phase are chosen such that the new component acquires a fit fraction of 5\%.
    \item $\Dp\Dm$ P~wave: Inclusion of the $\psi(4320)$ state, with fixed parameters~\cite{PDG2019}.
\end{itemize}
These effects related to the composition of the amplitude model constitute the largest systematic uncertainty for many of the parameters determined from the fit.

The statistical behaviour of the fit is investigated using pseudoexperiments, and the outcome of this study is used to correct the results of the fit to data as summarised in Table~\ref{tab:syst_summary}.
The model obtained from the best fit to data is used to generate an ensemble of datasets. Each dataset includes the efficiency variation across the Dalitz plot and a background contribution, the yield of which is sampled for each pseudoexperiment from a Poisson distribution centred at the observed background yield in data. Separate datasets are generated for Run~1 and Run~2 data. The standard fit is then applied to each dataset, where the signal yield is fixed to the generated value. 
Both the residual, $\left(P_{\rm fit} - P_{\rm gen}\right)$, and normalised residual or ``pull'', $\left(P_{\rm fit} - P_{\rm gen}\right)/\sigma_{\rm fit}$, are determined for the value $P$ of each parameter, determined with uncertainty $\sigma_{\rm fit}$, in the fit to each dataset.
The distribution of the residual for each fit parameter is fitted with a Gaussian function and the mean (``Bias'') is used to correct the central value.
The pull distribution for each fit parameter is also fitted with a Gaussian function, and the obtained width (``Pull width'') is used to scale the reported statistical uncertainty for the parameter.
For the fit fractions, which are calculated from the fitted complex coefficients, the width obtained from the fit of the distribution of the residuals with a Gaussian function is taken as the statistical uncertainty.

\begin{sidewaystable}
\caption{Systematic uncertainties on the complex coefficients and fit fractions of each component of the amplitude model: mass-fit signal shape (1), size of simulated sample for efficiencies (2), hardware trigger modelling (3), modelling: parent Blatt--Weisskopf radius (4), modelling: charmonia Blatt--Weisskopf radius (5), modelling: $\Dm\Kp$ resonances' Blatt--Weisskopf radius (6), model composition: S~wave (7), model composition: P~wave (8).\label{tab:syst_summary}}
\resizebox{\textwidth}{!}{
\begin{tabular}{l lr@{ $\pm$ }lllr@{ $\pm$ }llllllllll}
\hline\hline
\multicolumn{2}{c}{Parameter} & \multicolumn{2}{c}{Raw} & \multicolumn{1}{c}{Bias} & \multicolumn{1}{c}{Pull width} & \multicolumn{2}{c}{Corrected} & \multicolumn{1}{c}{(1)} & \multicolumn{1}{c}{(2)} & \multicolumn{1}{c}{(3)} & \multicolumn{1}{c}{(4)} & \multicolumn{1}{c}{(5)} & \multicolumn{1}{c}{(6)} & \multicolumn{1}{c}{(7)} & \multicolumn{1}{c}{(8)} & \multicolumn{1}{c}{(total)}\\
\hline
$\psi(3770)$ & Magnitude & \multicolumn{2}{c}{1 (fixed)} & \multicolumn{1}{l}{} & \multicolumn{1}{l}{} & \multicolumn{2}{l}{} & \multicolumn{1}{l}{} & \multicolumn{1}{l}{} & \multicolumn{1}{l}{} & \multicolumn{1}{l}{} & \multicolumn{1}{l}{} & \multicolumn{1}{l}{} & \multicolumn{1}{l}{} & \multicolumn{1}{l}{} & \multicolumn{1}{l}{}\\
 & Phase (rad) & \multicolumn{2}{c}{0 (fixed)} & \multicolumn{1}{l}{} & \multicolumn{1}{l}{} & \multicolumn{2}{l}{} & \multicolumn{1}{l}{} & \multicolumn{1}{l}{} & \multicolumn{1}{l}{} & \multicolumn{1}{l}{} & \multicolumn{1}{l}{} & \multicolumn{1}{l}{} & \multicolumn{1}{l}{} & \multicolumn{1}{l}{} & \multicolumn{1}{l}{}\\
 & Fit fraction & \multicolumn{2}{l}{\phm$0.144$} & $-0.001$ & \multicolumn{1}{l}{} & \phm$0.145$ & $0.012$ & \phm$0.001$ & \phm$0.001$ & \phm$0.002$ & \phm$0.000$ & \phm$0.006$ & \phm$0.000$ & \phm$0.001$ & \phm$0.004$ & \phm$0.008$\\
$\chi_{c0}(3930)$ & Magnitude & \phm$0.515$ & $0.068$ & \phm$0.009$ & \phm$0.945$ & \phm$0.506$ & $0.064$ & \phm$0.001$ & \phm$0.003$ & \phm$0.001$ & \phm$0.003$ & \phm$0.016$ & \phm$0.002$ & \phm$0.003$ & \phm$0.004$ & \phm$0.017$\\
 & Phase (rad) & \phm$2.168$ & $0.189$ & \phm$0.006$ & \phm$0.970$ & \phm$2.162$ & $0.184$ & \phm$0.000$ & \phm$0.011$ & \phm$0.003$ & \phm$0.003$ & \phm$0.027$ & \phm$0.009$ & \phm$0.011$ & \phm$0.012$ & \phm$0.034$\\
 & Fit fraction & \multicolumn{2}{l}{\phm$0.038$} & \phm$0.001$ & \multicolumn{1}{l}{} & \phm$0.037$ & $0.009$ & \phm$0.000$ & \phm$0.001$ & \phm$0.001$ & \phm$0.000$ & \phm$0.001$ & \phm$0.000$ & \phm$0.001$ & \phm$0.001$ & \phm$0.002$\\
$\chi_{c2}(3930)$ & Magnitude & \phm$0.704$ & $0.064$ & \phm$0.001$ & \phm$1.011$ & \phm$0.703$ & $0.064$ & \phm$0.001$ & \phm$0.003$ & \phm$0.005$ & \phm$0.004$ & \phm$0.006$ & \phm$0.001$ & \phm$0.006$ & \phm$0.003$ & \phm$0.012$\\
 & Phase (rad) & \phm$0.783$ & $0.170$ & $-0.044$ & \phm$1.003$ & \phm$0.827$ & $0.170$ & \phm$0.002$ & \phm$0.008$ & \phm$0.002$ & \phm$0.002$ & \phm$0.011$ & \phm$0.013$ & \phm$0.036$ & \phm$0.127$ & \phm$0.133$\\
 & Fit fraction & \multicolumn{2}{l}{\phm$0.072$} & $-0.001$ & \multicolumn{1}{l}{} & \phm$0.072$ & $0.012$ & \phm$0.001$ & \phm$0.001$ & \phm$0.000$ & \phm$0.001$ & \phm$0.002$ & \phm$0.000$ & \phm$0.000$ & \phm$0.003$ & \phm$0.003$\\
$\psi(4040)$ & Magnitude & \phm$0.608$ & $0.077$ & \phm$0.024$ & \phm$1.002$ & \phm$0.585$ & $0.078$ & \phm$0.003$ & \phm$0.005$ & \phm$0.003$ & \phm$0.001$ & \phm$0.021$ & \phm$0.007$ & \phm$0.009$ & \phm$0.024$ & \phm$0.035$\\
 & Phase (rad) & \phm$1.384$ & $0.188$ & $-0.032$ & \phm$0.938$ & \phm$1.416$ & $0.176$ & \phm$0.005$ & \phm$0.005$ & \phm$0.000$ & \phm$0.001$ & \phm$0.075$ & \phm$0.001$ & \phm$0.035$ & \phm$0.006$ & \phm$0.084$\\
 & Fit fraction & \multicolumn{2}{l}{\phm$0.053$} & \phm$0.003$ & \multicolumn{1}{l}{} & \phm$0.050$ & $0.013$ & \phm$0.000$ & \phm$0.001$ & \phm$0.000$ & \phm$0.000$ & \phm$0.002$ & \phm$0.001$ & \phm$0.002$ & \phm$0.003$ & \phm$0.004$\\
$\psi(4160)$ & Magnitude & \phm$0.670$ & $0.081$ & \phm$0.002$ & \phm$1.040$ & \phm$0.668$ & $0.084$ & \phm$0.003$ & \phm$0.005$ & \phm$0.004$ & \phm$0.000$ & \phm$0.014$ & \phm$0.004$ & \phm$0.015$ & \phm$0.047$ & \phm$0.052$\\
 & Phase (rad) & \phm$0.805$ & $0.232$ & $-0.093$ & \phm$0.970$ & \phm$0.898$ & $0.225$ & \phm$0.015$ & \phm$0.005$ & \phm$0.003$ & \phm$0.002$ & \phm$0.084$ & \phm$0.011$ & \phm$0.027$ & \phm$0.014$ & \phm$0.092$\\
 & Fit fraction & \multicolumn{2}{l}{\phm$0.065$} & $-0.001$ & \multicolumn{1}{l}{} & \phm$0.066$ & $0.015$ & \phm$0.000$ & \phm$0.000$ & \phm$0.000$ & \phm$0.000$ & \phm$0.000$ & \phm$0.001$ & \phm$0.001$ & \phm$0.012$ & \phm$0.012$\\
$\psi(4415)$ & Magnitude & \phm$0.770$ & $0.083$ & $-0.027$ & \phm$0.962$ & \phm$0.797$ & $0.080$ & \phm$0.000$ & \phm$0.005$ & \phm$0.009$ & \phm$0.005$ & \phm$0.011$ & \phm$0.015$ & \phm$0.053$ & \phm$0.022$ & \phm$0.061$\\
 & Phase (rad) & $-1.606$ & $0.229$ & $-0.148$ & \phm$0.860$ & $-1.458$ & $0.197$ & \phm$0.001$ & \phm$0.015$ & \phm$0.006$ & \phm$0.013$ & \phm$0.009$ & \phm$0.023$ & \phm$0.083$ & \phm$0.019$ & \phm$0.091$\\
 & Fit fraction & \multicolumn{2}{l}{\phm$0.086$} & $-0.006$ & \multicolumn{1}{l}{} & \phm$0.092$ & $0.014$ & \phm$0.000$ & \phm$0.001$ & \phm$0.001$ & \phm$0.001$ & \phm$0.006$ & \phm$0.004$ & \phm$0.011$ & \phm$0.008$ & \phm$0.015$\\
$X_0(2900)$ & Magnitude & \phm$0.628$ & $0.080$ & \phm$0.009$ & \phm$0.982$ & \phm$0.619$ & $0.079$ & \phm$0.003$ & \phm$0.006$ & \phm$0.001$ & \phm$0.001$ & \phm$0.001$ & \phm$0.002$ & \phm$0.005$ & \phm$0.024$ & \phm$0.025$\\
 & Phase (rad) & \phm$1.076$ & $0.201$ & $-0.015$ & \phm$0.961$ & \phm$1.091$ & $0.193$ & \phm$0.002$ & \phm$0.008$ & \phm$0.005$ & \phm$0.005$ & \phm$0.036$ & \phm$0.001$ & \phm$0.007$ & \phm$0.086$ & \phm$0.095$\\
 & Fit fraction & \multicolumn{2}{l}{\phm$0.057$} & \phm$0.001$ & \multicolumn{1}{l}{} & \phm$0.056$ & $0.014$ & \phm$0.000$ & \phm$0.001$ & \phm$0.001$ & \phm$0.000$ & \phm$0.002$ & \phm$0.000$ & \phm$0.003$ & \phm$0.003$ & \phm$0.005$\\
$X_1(2900)$ & Magnitude & \phm$1.432$ & $0.085$ & $-0.018$ & \phm$1.010$ & \phm$1.449$ & $0.086$ & \phm$0.005$ & \phm$0.010$ & \phm$0.010$ & \phm$0.002$ & \phm$0.012$ & \phm$0.001$ & \phm$0.011$ & \phm$0.023$ & \phm$0.032$\\
 & Phase (rad) & \phm$0.370$ & $0.109$ & \phm$0.003$ & \phm$0.939$ & \phm$0.367$ & $0.102$ & \phm$0.003$ & \phm$0.009$ & \phm$0.003$ & \phm$0.000$ & \phm$0.034$ & \phm$0.006$ & \phm$0.003$ & \phm$0.033$ & \phm$0.049$\\
 & Fit fraction & \multicolumn{2}{l}{\phm$0.296$} & $-0.010$ & \multicolumn{1}{l}{} & \phm$0.306$ & $0.024$ & \phm$0.000$ & \phm$0.002$ & \phm$0.000$ & \phm$0.001$ & \phm$0.007$ & \phm$0.000$ & \phm$0.006$ & \phm$0.019$ & \phm$0.021$\\
Nonres & Magnitude & \phm$1.301$ & $0.087$ & \phm$0.008$ & \phm$1.007$ & \phm$1.293$ & $0.088$ & \phm$0.006$ & \phm$0.009$ & \phm$0.010$ & \phm$0.003$ & \phm$0.029$ & \phm$0.003$ & \phm$0.002$ & \phm$0.029$ & \phm$0.043$\\
 & Phase (rad) & $-2.466$ & $0.125$ & $-0.056$ & \phm$0.953$ & $-2.410$ & $0.119$ & \phm$0.003$ & \phm$0.011$ & \phm$0.004$ & \phm$0.003$ & \phm$0.025$ & \phm$0.011$ & \phm$0.492$ & \phm$0.123$ & \phm$0.508$\\
 & Fit fraction & \multicolumn{2}{l}{\phm$0.244$} & \phm$0.002$ & \multicolumn{1}{l}{} & \phm$0.242$ & $0.022$ & \phm$0.001$ & \phm$0.001$ & \phm$0.001$ & \phm$0.001$ & \phm$0.001$ & \phm$0.000$ & \phm$0.001$ & \phm$0.004$ & \phm$0.005$\\
\hline\hline
\end{tabular}
}
\end{sidewaystable}

\begin{sidewaystable}
\caption{Systematic uncertainties on the masses (\nspgevcc) and widths (\nspgev) of the $\chi_{c0,2}(3930)$ and $X_{0,1}(2900)$ resonances: mass-fit signal shape (1), size of simulated sample for efficiencies (2), hardware trigger modelling (3), modelling: parent Blatt--Weisskopf radius (4), modelling: charmonia Blatt--Weisskopf radius (5), modelling: $\Dm\Kp$ resonances' Blatt--Weisskopf radius (6), model composition: S~wave (7), model composition: P~wave (8).\label{tab:syst_shapeParams_summary}}
\resizebox{\textwidth}{!}{
\begin{tabular}{l lr@{ $\pm$ }lllr@{ $\pm$ }llllllllll}
\hline\hline
\multicolumn{2}{l}{Parameter} & \multicolumn{2}{c}{Raw} & \multicolumn{1}{c}{Bias} & \multicolumn{1}{c}{Pull width} & \multicolumn{2}{c}{Corrected} & \multicolumn{1}{c}{(1)} & \multicolumn{1}{c}{(2)} & \multicolumn{1}{c}{(3)} & \multicolumn{1}{c}{(4)} & \multicolumn{1}{c}{(5)} & \multicolumn{1}{c}{(6)} & \multicolumn{1}{c}{(7)} & \multicolumn{1}{c}{(8)} & \multicolumn{1}{c}{(total)}\\
\hline
$\chi_{c0}(3930)$ & Mass & \phm$3.9239$ & $0.0015$ & \phm$0.0002$ & \phm$1.0004$ & \phm$3.9238$ & $0.0015$ & \phm$0.0000$ & \phm$0.0000$ & \phm$0.0000$ & \phm$0.0000$ & \phm$0.0000$ & \phm$0.0000$ & \phm$0.0001$ & \phm$0.0004$ & \phm$0.0004$\\
          & Width & \phm$0.0170$ & $0.0050$ & $-0.0004$ & \phm$1.0255$ & \phm$0.0174$ & $0.0051$ & \phm$0.0001$ & \phm$0.0002$ & \phm$0.0002$ & \phm$0.0001$ & \phm$0.0005$ & \phm$0.0000$ & \phm$0.0003$ & \phm$0.0003$ & \phm$0.0008$\\
$\chi_{c2}(3930)$ & Mass & \phm$3.9262$ & $0.0023$ & $-0.0006$ & \phm$1.0500$ & \phm$3.9268$ & $0.0024$ & \phm$0.0000$ & \phm$0.0000$ & \phm$0.0000$ & \phm$0.0000$ & \phm$0.0002$ & \phm$0.0000$ & \phm$0.0004$ & \phm$0.0006$ & \phm$0.0008$\\
          & Width & \phm$0.0337$ & $0.0064$ & $-0.0005$ & \phm$1.0287$ & \phm$0.0342$ & $0.0066$ & \phm$0.0000$ & \phm$0.0002$ & \phm$0.0002$ & \phm$0.0003$ & \phm$0.0005$ & \phm$0.0001$ & \phm$0.0003$ & \phm$0.0009$ & \phm$0.0011$\\
$X_0(2900)$ & Mass & \phm$2.8670$ & $0.0063$ & \phm$0.0007$ & \phm$1.0262$ & \phm$2.8663$ & $0.0065$ & \phm$0.0000$ & \phm$0.0002$ & \phm$0.0000$ & \phm$0.0002$ & \phm$0.0003$ & \phm$0.0002$ & \phm$0.0002$ & \phm$0.0019$ & \phm$0.0020$\\
          & Width & \phm$0.0570$ & $0.0121$ & $-0.0003$ & \phm$1.0123$ & \phm$0.0572$ & $0.0122$ & \phm$0.0006$ & \phm$0.0004$ & \phm$0.0004$ & \phm$0.0001$ & \phm$0.0007$ & \phm$0.0005$ & \phm$0.0007$ & \phm$0.0038$ & \phm$0.0041$\\
$X_1(2900)$ & Mass & \phm$2.9053$ & $0.0050$ & \phm$0.0011$ & \phm$0.9629$ & \phm$2.9041$ & $0.0048$ & \phm$0.0002$ & \phm$0.0001$ & \phm$0.0002$ & \phm$0.0001$ & \phm$0.0007$ & \phm$0.0003$ & \phm$0.0004$ & \phm$0.0009$ & \phm$0.0013$\\
          & Width & \phm$0.1088$ & $0.0105$ & $-0.0015$ & \phm$1.0154$ & \phm$0.1103$ & $0.0107$ & \phm$0.0008$ & \phm$0.0003$ & \phm$0.0004$ & \phm$0.0002$ & \phm$0.0012$ & \phm$0.0010$ & \phm$0.0001$ & \phm$0.0038$ & \phm$0.0043$\\
\hline\hline
\end{tabular}
}
\end{sidewaystable}

\section{Significance of resonant structures}
\label{sec_significanceTests}
\begin{figure}[!tb]
\centering
    \includegraphics[width=.45\textwidth]{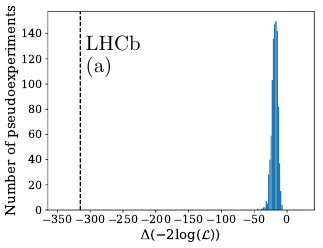}
    \includegraphics[width=.45\textwidth]{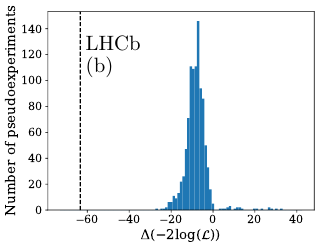}\\ \includegraphics[width=.45\textwidth]{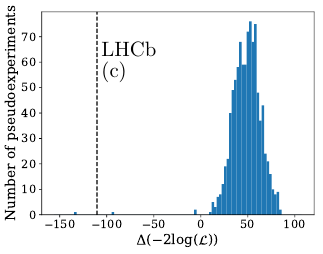}
    \includegraphics[width=.45\textwidth]{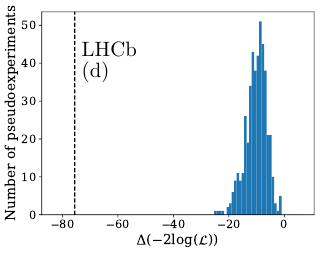}            
    \caption{Distributions of the test-statistic $t$ in ensembles of pseudoexperiments generated according to  various hypotheses and compared to values found in data (indicated by dashed vertical lines). In (a), the $H_0$ hypothesis is a model fit to data without $\Dm\Kp$ resonances. In (b), (c) and (d) plots, the $H_0$ hypothesis assumes a single $\chi_{cJ}(3930)$ state, which has spin-0, spin-1 and spin-2, respectively. 
    \label{im_signifTests_signifGrid}}
\end{figure}

Pseudoexperiments are used to determine the significance of the $\Dm\Kp$ structure.
The pseudoexperiments are generated using an amplitude model where no $\Dm\Kp$ resonances are included, with parameters obtained by fitting the data (see Sec.~\ref{sec:results_noDK}). For each dataset, the yields of the signal and background components are sampled from a Poisson distribution centred at the yields observed in the data, and the efficiency is applied to the signal component. 
Each dataset is fitted with both the model used for generation ($H_0$) and the baseline fit model ($H_1$) and the test statistic $t=-2(\log(\mathcal{L}(H_1) - \log(\mathcal{L}(H_0)))$ is determined.
The test statistic observed in data is compared to the distribution from the pseudoexperiments in Fig.~\ref{im_signifTests_signifGrid}(a), where the preference for the nominal hypothesis is overwhelming.
These results confirm those of Sec.~\ref{sec:other_models}.

The significance of the $X_1(2900)$ and $X_0(2900)$ states in this amplitude analysis is much larger than the significance of exotic contributions obtained in the model-independent analysis of the same data sample~\cite{LHCb-PAPER-2020-024}.
This is expected since in the model-independent analysis the contributions from S, P and D waves in the \Dp\Dm system are independent in each $m(\Dp\Dm)$ bin, while in the amplitude analysis each partial wave is a continuous function of $m(\Dp\Dm)$ that is prescribed by the model.
The amplitude analysis consequently has less freedom to absorb any structure in the $m(\Dm\Kp)$ distribution compared to the model-independent approach, unless explicit components are included to describe it, and correspondingly a higher significance is obtained.

A similar approach is taken to determine the significance of the presence of both spin-0 and spin-2 states in the $\chi_{cJ}(3930)$ region. Three alternative configurations are considered, where these two components are replaced by a single resonance, having spin 0, 1, or 2. 
The results are shown in Fig.~\ref{im_signifTests_signifGrid}.
The smallest, though still compelling, significance of the two state fit occurs when comparing to a single spin-1 resonance in the $\chi_{cJ}(3930)$ region.
Hence the need for two states in this region is clearly established.
These results also confirm those of Sec.~\ref{sec:other_models}, where issues with fits including a spin-1 state in the $\chi_{cJ}(3930)$ region are discussed, leaving the configuration with spins 0 and 2 as the only candidate to describe the data.

\section{Summary}
\label{sec_summary}

The first amplitude analysis of the \BpDpDmKp\ decay has been carried out. 
The analysis is performed using \lhcb $pp$ collision data taken at $\sqrt{s}=7,8,$ and $13\tev$, corresponding to a total integrated luminosity of 9\invfb, from which a highly pure sample of 1260 signal candidates are selected.

It is not possible to describe the distribution across the Dalitz plot using only resonances in the $\Dp\Dm$ system; this conclusion is supported by a model-independent analysis of the same data sample~\cite{LHCb-PAPER-2020-024}. 
Reasonable agreement with the data is achieved by including new spin-0 and spin-1 resonances in the $\Dm\Kp$ channel, described with Breit--Wigner lineshapes, the parameters of which are determined to be 
\begin{eqnarray*}
X_0(2900): \quad 
M = \round{3}{2.8663} \pm \round{3}{0.0065} \pm \round{3}{0.0020} \gevcc\,, & &  
\Gamma = \phz\round{0}{57.2} \pm \round{0}{12.2} \pm \round{0}{4.1} \mev \,, \\
X_1(2900): \quad
M = \round{3}{2.9041} \pm \round{3}{0.0048} \pm \round{3}{0.0013} \gevcc\,, & & 
\Gamma = \round{0}{110.3} \pm \round{0}{10.7} \pm \round{0}{4.3} \mev \,,
\end{eqnarray*}
where the first uncertainties are statistical and the second systematic.
While the significance of these contributions is overwhelming, and this model gives a good description of the data in this region, it cannot be ruled out that alternative models incorporating additional hadronic effects such as rescattering may also be able to accommodate these $\Dm\Kp$ structures.  
Nonetheless, if the $\Dm\Kp$ structures are interpreted as resonances, these results constitute the first clear observation of exotic hadrons with open flavour, and the first that do not contain a heavy quark-antiquark pair.
More detailed investigations will require larger data samples and studies of additional decay modes.  
For example, it will be interesting to see if similar structures can be observed in $\Bp \to \Dm\Kp\pip$ decays, where an analysis of a subset of the existing LHCb data sample~\cite{LHCb-PAPER-2015-007} gave an indication of an excess --- though not statistically significant --- in the $m(\Dm\Kp)$ region where structure is now observed.

The model also includes contributions from the $\psi(3770)$, $\psi(4040)$, $\psi(4160)$ and $\psi(4415)$ vector charmonia states.
In addition, it is found necessary to include both spin-0 and spin-2 states in the $\chi_{cJ}(3930)$ region, the parameters of which are determined from the fit to be
\begin{eqnarray*}
\chi_{c0}(3930): \quad
M = 3.9238 \pm 0.0015 \pm 0.0004 \gevcc \,, & &
\Gamma = 17.4 \pm 5.1 \pm 0.8 \mev\,,\\
\chi_{c2}(3930): \quad
M = 3.9268 \pm 0.0024 \pm 0.0008 \gevcc\,, & &
\Gamma = 34.2 \pm 6.6 \pm 1.1 \mev\,.
\end{eqnarray*}
Previous measurements of the properties of the $\chi_{c2}(3930)$ state have assumed a single state in this region and, in the light of these results, may be unreliable.
There is no evidence for the $\chi_{c0}(3860)$ state reported by the Belle collaboration~\cite{Chilikin:2017evr}.
Further investigation and independent confirmation of these results concerning spin-0 and spin-2 charmonium states may be obtained in future by studies of $\Bp \to \jpsi \omega \Kp$ decays.

The size and purity of the sample demonstrates the potential impact of further studies of $B \to D\Dbar K$ decays in the LHCb dataset.  
In particular, the $\Bp \to \Dz\Dzb\Kp$ mode is likely to shed further light on the production of charmonium states in $B$-meson decays, while analysis of $\Bz \to \Dz\Dm\Kp$ may provide crucial additional information on the $\Dm\Kp$ structures. 
In both cases, however, contributions from $\Ds$ excitations decaying to $\Dz\Kp$ will also need to be considered.
The significantly larger sample anticipated to be collected by LHCb with an upgraded detector during Run~3 of the Large Hadron Collider also provides exciting prospects for further discoveries in this area.

%% file: acknowledgements.tex
\section*{Acknowledgements}
%
%
\noindent We express our gratitude to our colleagues in the CERN
accelerator departments for the excellent performance of the LHC. We
thank the technical and administrative staff at the LHCb
institutes.
We acknowledge support from CERN and from the national agencies:
CAPES, CNPq, FAPERJ and FINEP (Brazil); 
MOST and NSFC (China); 
CNRS/IN2P3 (France); 
BMBF, DFG and MPG (Germany); 
INFN (Italy); 
NWO (Netherlands); 
MNiSW and NCN (Poland); 
MEN/IFA (Romania); 
MSHE (Russia); 
MICINN (Spain); 
SNSF and SER (Switzerland); 
NASU (Ukraine); 
STFC (United Kingdom); 
DOE NP and NSF (USA).
We acknowledge the computing resources that are provided by CERN, IN2P3
(France), KIT and DESY (Germany), INFN (Italy), SURF (Netherlands),
PIC (Spain), GridPP (United Kingdom), RRCKI and Yandex
LLC (Russia), CSCS (Switzerland), IFIN-HH (Romania), CBPF (Brazil),
PL-GRID (Poland) and OSC (USA).
We are indebted to the communities behind the multiple open-source
software packages on which we depend.
Individual groups or members have received support from
AvH Foundation (Germany);
EPLANET, Marie Sk\l{}odowska-Curie Actions and ERC (European Union);
A*MIDEX, ANR, Labex P2IO and OCEVU, and R\'{e}gion Auvergne-Rh\^{o}ne-Alpes (France);
Key Research Program of Frontier Sciences of CAS, CAS PIFI,
Thousand Talents Program, and Sci. \& Tech. Program of Guangzhou (China);
RFBR, RSF and Yandex LLC (Russia);
GVA, XuntaGal and GENCAT (Spain);
the Royal Society
and the Leverhulme Trust (United Kingdom).

%% file: appendix.tex

\clearpage
\section*{Appendices}

\appendix
\section{Helicity-angle distributions in slices of Dalitz-plot variables}
\label{app:helicity}

\begin{figure}[htbp]
\centering
    \includegraphics[width=.49\textwidth]{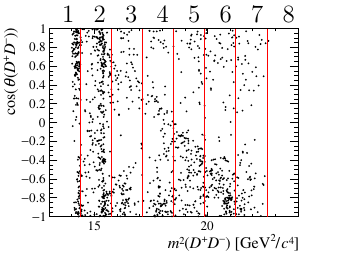}
    \includegraphics[width=.49\textwidth]{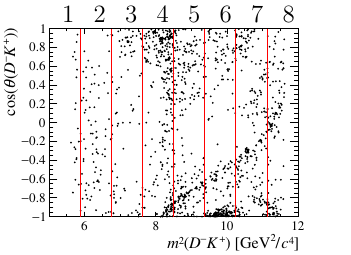}\\
    \includegraphics[width=.49\textwidth]{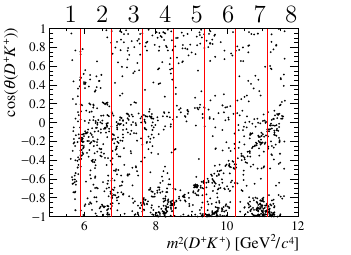} 
\caption{Division of the square Dalitz plot in slices of invariant mass squared. The binning is used for (top left) the $\cos\left(\theta(\Dp\Dm)\right)$ distribution, (top right) the $\cos\left(\theta(\Dm\Kp)\right)$ distribution, and (lower) the $\cos\left(\theta(\Dp\Kp)\right)$ distribution. 
\label{im:sqDalBinning}}
\end{figure}

\begin{figure}[htbp]
\centering
    \includegraphics[width=.32\textwidth]{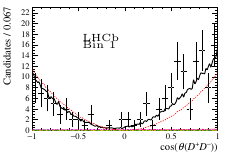}
    \includegraphics[width=.32\textwidth]{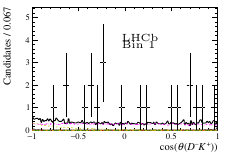}
    \includegraphics[width=.32\textwidth]{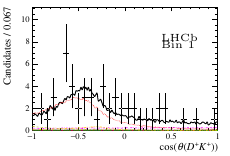}\\
    \includegraphics[width=.32\textwidth]{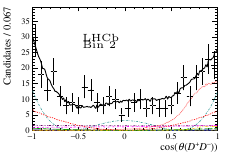}
    \includegraphics[width=.32\textwidth]{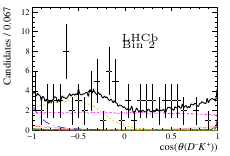}
    \includegraphics[width=.32\textwidth]{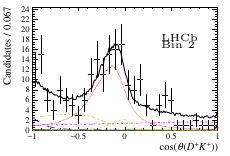}\\
    \includegraphics[width=.32\textwidth]{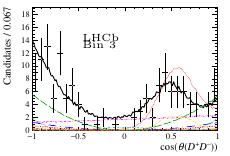}
    \includegraphics[width=.32\textwidth]{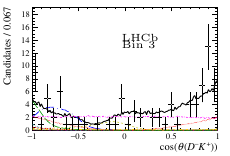}
    \includegraphics[width=.32\textwidth]{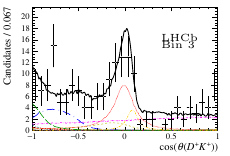}\\    
    \includegraphics[width=.32\textwidth]{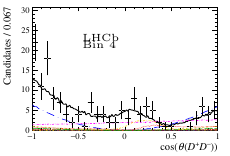}
    \includegraphics[width=.32\textwidth]{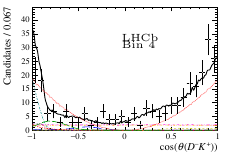}
    \includegraphics[width=.32\textwidth]{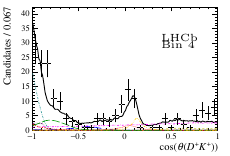}
\caption{Helicity-angle distributions divided according to the binning scheme shown in Fig.~\ref{im:sqDalBinning} (bins 1-4). The different components are shown as indicated in the legend of Fig.~\ref{im_model1_BmDpDmKm}.\label{im_model1_BmDpDmKm_helicity_both_1to4}}
\end{figure}

\begin{figure}[htbp]
\centering
    \includegraphics[width=.32\textwidth]{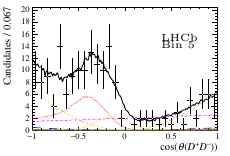}
    \includegraphics[width=.32\textwidth]{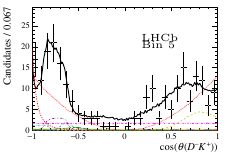}
    \includegraphics[width=.32\textwidth]{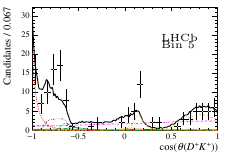}\\
    \includegraphics[width=.32\textwidth]{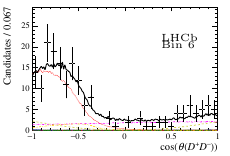}
    \includegraphics[width=.32\textwidth]{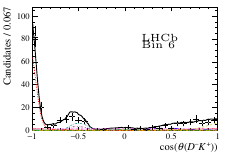}
    \includegraphics[width=.32\textwidth]{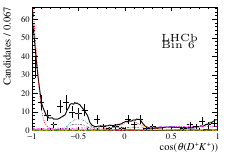}\\
    \includegraphics[width=.32\textwidth]{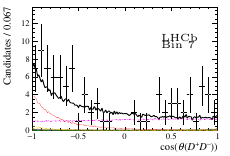}
    \includegraphics[width=.32\textwidth]{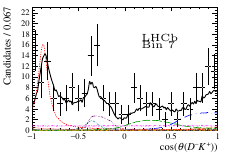}
    \includegraphics[width=.32\textwidth]{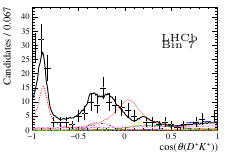}\\    
    \includegraphics[width=.32\textwidth]{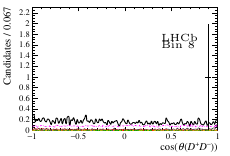}
    \includegraphics[width=.32\textwidth]{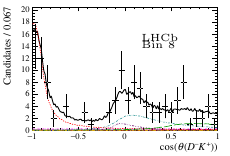}
    \includegraphics[width=.32\textwidth]{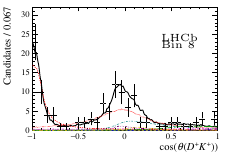}
\caption{Helicity-angle distributions divided according to the binning scheme shown in Fig.~\ref{im:sqDalBinning} (bins 5-8). The different components are shown as indicated in the legend of Fig.~\ref{im_model1_BmDpDmKm}.\label{im_model1_BmDpDmKm_helicity_both_5to8}}
\end{figure}

To allow detailed inspection of the agreement between the result of the fit and the data, helicity angle distributions are shown in slices of the three invariant mass-squared combinations.
Figure~\ref{im:sqDalBinning} defines the slices for these projections, with the helicity angle distributions themselves shown in Figs.~\ref{im_model1_BmDpDmKm_helicity_both_1to4}--\ref{im_model1_BmDpDmKm_helicity_both_5to8}.

\clearpage
\section{Angular moments}
\label{app:moments}

\begin{figure}[htbp]
\centering
    \includegraphics[width=.32\textwidth]{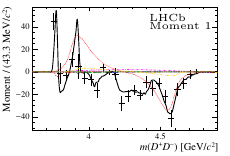}
    \includegraphics[width=.32\textwidth]{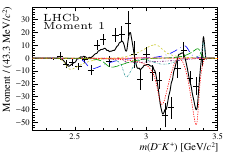}
    \includegraphics[width=.32\textwidth]{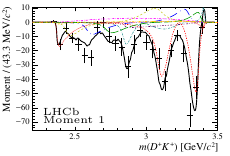}\\
    \includegraphics[width=.32\textwidth]{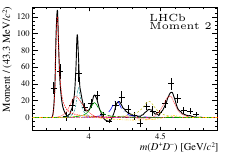}
    \includegraphics[width=.32\textwidth]{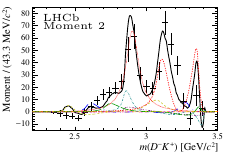}
    \includegraphics[width=.32\textwidth]{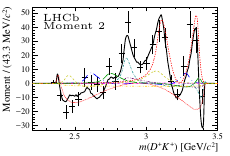}\\
    \includegraphics[width=.32\textwidth]{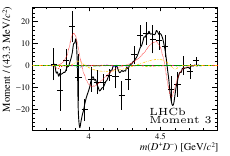}
    \includegraphics[width=.32\textwidth]{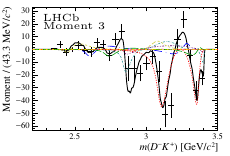}
    \includegraphics[width=.32\textwidth]{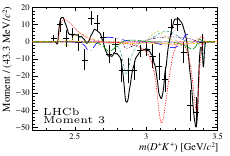}\\    
    \includegraphics[width=.32\textwidth]{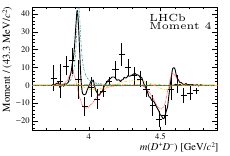}
    \includegraphics[width=.32\textwidth]{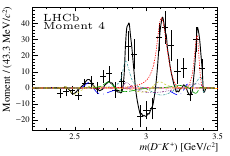}
    \includegraphics[width=.32\textwidth]{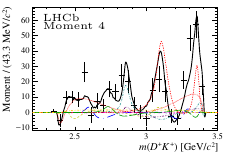}\\
    \includegraphics[width=.32\textwidth]{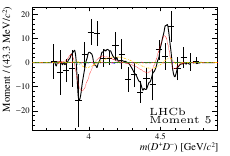}
    \includegraphics[width=.32\textwidth]{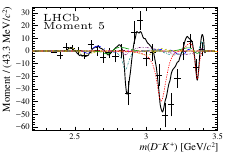}
    \includegraphics[width=.32\textwidth]{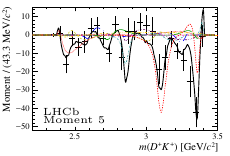}    
\caption{Projections of moments 1--5 of each pair of final-state particles in the \BpDpDmKp Dalitz plot. As usual, data points are shown in black and the total, and individual components' PDFs are overlaid. The different components are shown as indicated in the legend of Fig.~\ref{im_model1_BmDpDmKm}. \label{im_model1_BmDpDmKm_moments1to5_both}}
\end{figure}

\begin{figure}[htbp]
 \centering
    \includegraphics[width=.32\textwidth]{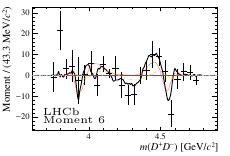}
    \includegraphics[width=.32\textwidth]{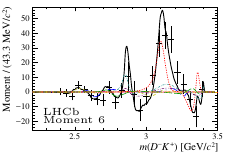}
    \includegraphics[width=.32\textwidth]{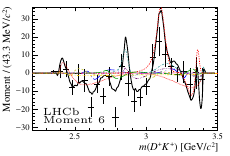}\\
    \includegraphics[width=.32\textwidth]{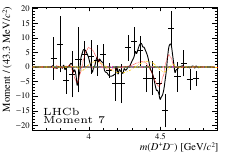}
    \includegraphics[width=.32\textwidth]{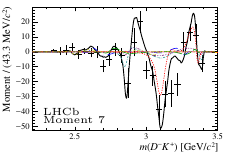}
    \includegraphics[width=.32\textwidth]{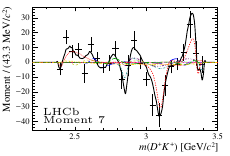}\\
    \includegraphics[width=.32\textwidth]{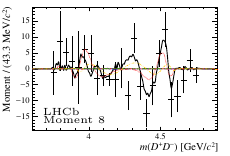}
    \includegraphics[width=.32\textwidth]{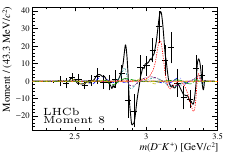}
    \includegraphics[width=.32\textwidth]{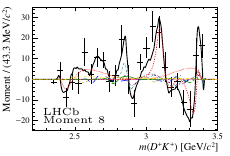}\\    
    \includegraphics[width=.32\textwidth]{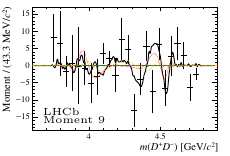}
    \includegraphics[width=.32\textwidth]{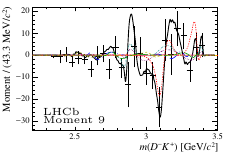}
    \includegraphics[width=.32\textwidth]{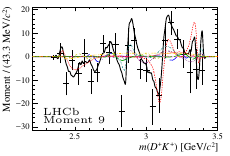}
\caption{Projections of moments 6--9 of each pair of final-state particles in the \BpDpDmKp Dalitz plot. As usual, data points are shown in black and the total, and individual components' PDFs are overlaid. The different components are shown as indicated in the legend of Fig.~\ref{im_model1_BmDpDmKm}. \label{im_model1_BmDpDmKm_moments6to9_both}}
 \end{figure}

The angular moments of the data, in bins of $m(\Dp\Dm)$, are central to the model-independent analysis presented in Ref.~\cite{LHCb-PAPER-2020-024}.
They also present a further way of checking the agreement between the result of the fit and the data.
Moments 1--5, for each of $m(\Dp\Dm)$, $m(\Dm\Kp)$ and $m(\Dp\Kp)$ are presented in Fig.~\ref{im_model1_BmDpDmKm_moments1to5_both}, with moments 6--9 in Fig.~\ref{im_model1_BmDpDmKm_moments6to9_both}.

\clearpage

%% file: LHCb_Authorship_21-Jul-2020.tex
\centerline
{\large\bf LHCb collaboration}
\begin
{flushleft}
\small
R.~Aaij$^{31}$,
C.~Abell{\'a}n~Beteta$^{49}$,
T.~Ackernley$^{59}$,
B.~Adeva$^{45}$,
M.~Adinolfi$^{53}$,
H.~Afsharnia$^{9}$,
C.A.~Aidala$^{84}$,
S.~Aiola$^{25}$,
Z.~Ajaltouni$^{9}$,
S.~Akar$^{64}$
% ...

{\footnotesize \it

% ...
$ ^{9}$Universit{\'e} Clermont Auvergne, CNRS/IN2P3, LPC, Clermont-Ferrand, France\\
% ...
$ ^{25}$INFN Sezione di Milano, Milano, Italy\\
% ...
$ ^{31}$Nikhef National Institute for Subatomic Physics, Amsterdam, Netherlands\\
% ...
$ ^{45}$Instituto Galego de F{\'\i}sica de Altas Enerx{\'\i}as (IGFAE), Universidade de Santiago de Compostela, Santiago de Compostela, Spain\\
% ...
$ ^{49}$Physik-Institut, Universit{\"a}t Z{\"u}rich, Z{\"u}rich, Switzerland\\
% ...
$ ^{53}$H.H. Wills Physics Laboratory, University of Bristol, Bristol, United Kingdom\\
% ...
$ ^{59}$Oliver Lodge Laboratory, University of Liverpool, Liverpool, United Kingdom\\
% ...
$ ^{64}$University of Cincinnati, Cincinnati, OH, United States\\
% ...
$ ^{67}$Syracuse University, Syracuse, NY, United States\\
% ...
$ ^{84}$University of Michigan, Ann Arbor, United States, associated to $^{67}$\\
\bigskip
% ...
\medskip
}
\end{flushleft}